\documentclass[twocolumn]{aastex63}

\usepackage{ulem}
\usepackage{booktabs}
\usepackage{multirow}
\usepackage{xcolor}
\usepackage{bm}

\shorttitle{Cepheid Distances from EDR3}
\shortauthors{Owens et al.}

\graphicspath{{./}{figures/}}

\begin{document}

\title{Current Challenges in Cepheid Distance Calibrations Using \textit{Gaia} EDR3}

\author{Kayla~A.~Owens}\affil{Department of Astronomy \& Astrophysics, University of Chicago, 5640 South Ellis Avenue, Chicago, IL 60637}\affiliation{Kavli Institute for Cosmological Physics, University of Chicago, 5640 South Ellis Avenue, Chicago, IL 60637}

\author{Wendy~L.~Freedman}\affil{Department of Astronomy \& Astrophysics, University of Chicago, 5640 South Ellis Avenue, Chicago, IL 60637}\affiliation{Kavli Institute for Cosmological Physics, University of Chicago, 5640 South Ellis Avenue, Chicago, IL 60637}

\author{Barry~F.~Madore}\affil{Observatories of the Carnegie Institution for Science 813 Santa Barbara St., Pasadena, CA~91101}\affil{Department of Astronomy \& Astrophysics, University of Chicago, 5640 South Ellis Avenue, Chicago, IL 60637}

\author{Abigail~J.~Lee}\affil{Department of Astronomy \& Astrophysics, University of Chicago, 5640 South Ellis Avenue, Chicago, IL 60637}\affiliation{Kavli Institute for Cosmological Physics, University of Chicago, 5640 South Ellis Avenue, Chicago, IL 60637}
\correspondingauthor{Kayla~A.~Owens}\email{kaowens@uchicago.edu}

\begin{abstract}

Using parallaxes from \textit{Gaia} Early Data Release 3 (EDR3), we determine multi-wavelength $BVI_{c}$, $JHK_{s}$ and $[3.6]$ \& $[4.5]$ micron absolute magnitudes for 37 nearby Milky Way Cepheids, covering the period range between 5 and 60 days. We apply these period-luminosity relations to Cepheids in the Large and Small Magellanic Clouds, and find that the derived distances are significantly discrepant with the geometric distances according to detached eclipsing binaries (DEBs). We explore several potential causes of these issues, including reddening, metallicity, and the existence of an additional zero-point offset, but none provide a sufficient reconciliation with both DEB distances. We conclude that the combination of the systematic uncertainties on the EDR3 parallaxes with the uncertainties on the effect of metallicity on the Cepheid distance scale leads to a systematic error floor of approximately $3$\%. We therefore find that the EDR3 data is not sufficiently accurate in the regime of these bright Cepheids to determine extragalactic distances precise to the 1\% level at this time, in agreement with a number of contemporary studies.

\end{abstract}


\section{Introduction}\label{sec:intro}

Parallax-based calibrations of the Cepheid period-luminosity (PL) relation (or Leavitt Law) have become increasingly precise in the past two decades. The \textit{Hipparcos} mission provided the first high-precision stellar parallaxes in 1997 \citep{Perryman1997}, which were quickly applied to measurements of the Cepheid distance scale by \cite{1997MNRAS.286L...1F}. However, the zero-point calibration of these parallaxes was uncertain, and final values depended heavily upon adopted slopes and sample selection \citep{1998ApJ...492..110M}. Later, \cite{2002AJ....124.1695B, 2007AJ....133.1810B} used the Fine Guidance Sensor on the Hubble Space Telescope (HST) to determine individual trigonometric parallaxes to 10 Galactic Cepheid variables having an average precision of 8\%. This parallax sample has been widely used to calibrate the absolute PL relations at many wavelengths and then to calculate distances to other galaxies, as for example by \cite{2007AA...476...73F, 2007MNRAS.379..723V, 2012ApJ...759..146M}. More recently, \cite{2014ApJ...785..161R} used the spatial scanning mode of HST to measure the parallax to the long-period Galactic Cepheid SY Aur. \cite{2018ApJ...855..136R} then applied this same methodology to measure parallaxes to 7 more long-period Galactic Cepheids, yielding a total sample of 8 HST scanning mode Cepheids with an average parallax error of 12\%.

The recent \textit{Gaia} Early Data Release 3 (EDR3) has provided parallaxes to nearly 1.5 billion stars \citep{2020arXiv201201533G}. This gives us an opportunity to calibrate the absolute Cepheid PL relation, based on a larger and more distant sample of Cepheids than was possible with either \textit{Hipparcos} or HST. 

Precisely calibrating the Cepheid PL relation is of paramount importance to cosmology, given its common application to the extragalactic distance scale. Cepheid distances are used as a means of calibrating type-Ia supernovae to measure the Hubble constant ($H_{0}$), as for example by the HST Key Project \citep{2001ApJ...553...47F} and subsequently by the SH0ES Project \citep{2016ApJ...826...56R, 2019ApJ...876...85R}. Uncertainties in distances to the closest galaxies propagate to further Cepheid-based distance measurements, and in an era ambitiously aiming for accuracies of 1\% in $H_{0}$ \citep{2021ApJ...908L...6R}, minimizing these uncertainties is essential. Clearly, \textit{Gaia}'s potential to measure extremely precise absolute distances will eventually be instrumental in calibrating the local distance scale at the accuracy and precision necessary.

The formal errors on the \textit{Gaia} EDR3 parallaxes are modestly underestimated, according to recent publications. For example, \cite{2020arXiv201206242F} investigated the completeness, accuracy, and precision of the EDR3 catalog. They calculate the unit weight uncertainties of the catalog, which are the factors by which the formal errors must be increased to represent the true level of uncertainty. The multiplicative factor is only around 1.2 for most stars, although it can rise to a factor of more than 2 in the worst cases. Further, the errors have been shown to be most significantly underestimated for brighter stars \citep{2021MNRAS.tmp..394E}. Unfortunately, many of the Milky Way field Cepheids used for distance determinations lie in this very bright apparent magnitude range, with our own sample having magnitudes between $4<G<11$ mag. Additionally, Cepheids vary dramatically in temperature over their cycle, averaging about 1000K from maximum to minimum light \citep{2018A&A...616A..82P}, corresponding to a peak-to-peak color variation of about 0.5~mag in the optical region. Since the Gaia parallax offset is known to vary with magnitude and color, this could introduce larger parallax errors for individual Cepheids. \cite{2020arXiv201203380L} state that the pipeline for variable stars has significantly improved since DR2; however, since the color of a star is still assumed to be the same in all observations, corrections for the chromaticity may not be fully accurate, so there are still issues to be solved in future data releases.

Only a few EDR3 studies using primarily very bright sources have been released, but several of them have speculated on the existence of a distinct zero-point offset for sources with $G\lesssim11$ mag \citep[e.g.,][]{2021ApJ...908L...6R, 2021arXiv210107252Z, 2021arXiv210109691H}, though the significance of the offset is disputed \citep{2021ApJ...907L..33S}. Furthermore, the existence of such offsets has been shown to be degenerate with the derived metallicity effects on the PL relations \citep{2021arxiv210811391R}. With recent literature disagreeing on the exact values of these effects across photometric bands, we feel that an accurate calibration of the Cepheid distance scale via Milky Way Cepheids is not yet within reach. Because of these numerous difficulties, rather than presenting a single calibration, we explore the effects that small differences in analysis choices can have on the final results of distance calibrations to the LMC and SMC, with the aim of better understanding the current uncertainties.

In Section \ref{sec:phot}, we describe our adopted multi-wavelength photometric samples for the Milky Way, the LMC, and the SMC. In Section \ref{sec:MWdist}, we describe determining distances to individual Milky Way Cepheids and to the LMC and SMC. In Section \ref{sec:Qual}, we investigate the overall quality of the EDR3 data for bright Cepheids and compare our set of measurements to parallaxes from HST and prior distance measurements to the Magellanic Clouds based on detached eclipsing binaries (DEBs). Finally, in Section \ref{sec:err_bud}, we calculate the statistical and systematic errors on these measurements.
\medskip

\section{Photometric Samples}\label{sec:phot}

In Sections \ref{subsec:MWphot}, \ref{subsec:LMCphot}, and \ref{subsec:SMCphot} we describe collecting multi-wavelength $BVI_{c}$, $JHK_{s}$ and $[3.6]$ \& $[4.5]$ micron data from the literature for the Milky Way, LMC, and SMC, respectively. We then describe our adopted period cuts and the removal of low-quality Milky Way Cepheids in Section \ref{subsec:refphot}. 

\subsection{Milky Way}\label{subsec:MWphot}
To calibrate the absolute PL relation in the Milky Way, we use multi-wavelength $BVI_{c}$ and $JHK_{s}$ data for 59 Galactic Cepheids from \cite{2007AA...476...73F}. Their optical sample ($BVI_{c}$) was compiled directly from the literature, primarily from the catalog of \cite{2000A&AS..143..211B} which gives mean $BVRI$ and $R_{c}I_{c}$ magnitudes for 455 Galactic Cepheids. Their near-IR sample ($JHK_{s}$) was generated by combining intensity-mean values from \cite{1984ApJS...54..547W, 1992A&AS...93...93L, 2003ApJ...592..539B} and converting them to the 2MASS photometric system \citep{2006AJ....131.1163S}. We obtained $[3.6]$ and $[4.5]$ micron data for the 29 Cepheids in common with \cite{2012ApJ...759..146M}, who collected ``warm" \textit{Spitzer} observations (taken in the post-cryogenic part of the extended mission) for 37 well-observed Galactic Cepheids used to calibrate the mid-IR PL relations. These observations were scheduled to evenly sample the phase space of each Cepheid at 24 epochs, allowing for very accurate determinations of their mean magnitudes. This even-phase sampling was performed for all of the mid-IR observation sets, including the LMC and SMC samples described below. By design, the Milky Way sample itself evenly populates the period range between 5 and 45 days. The Milky Way Cepheid periods, extinctions, parallaxes, distance moduli, and photometry are all compiled in Table \ref{tab:ceph_phot}.

\begin{longrotatetable}
\begin{deluxetable*}{lccclrrrrrrrrccccc}
\tabletypesize{\footnotesize}
\tablecaption{Cepheid Properties and Photometry}\label{tab:ceph_phot}

\tablehead{\colhead{Cepheid} & \colhead{$\log P$} & \colhead{$E(B-V)$} & \colhead{$\pi_{\mathit{Gaia}}$} & \colhead{$\pi_{\mathrm{Fouque}}$} & \colhead{$(m - M)_0^{\mathit{Gaia}}$} & \colhead{$(m - M)_0^{\mathrm{Fouque}}$} & \colhead{$B$} & \colhead{$V$} & \colhead{$I_{c}$} & \colhead{$J$} & \colhead{$H$} & \colhead{$K_{s}$} & \colhead{$[3.6]$} & \colhead{$[4.5]$} & \colhead{{\tt ruwe}} \\ 
\colhead{} & \colhead{(days)} & \colhead{(mag)} & \colhead{(mas)} & \colhead{(mas)} & \colhead{(mag)} & \colhead{(mag)} & \colhead{(mag)} & \colhead{(mag)} & \colhead{(mag)} & \colhead{(mag)} & \colhead{(mag)} & \colhead{(mag)} & \colhead{(mag)} & \colhead{(mag)} & \colhead{}} 

\startdata
RT Aur \tablenotemark{\scriptsize a}\tablenotemark{\scriptsize b} &0.571&0.062& 1.858$\pm$0.122 & 2.40$\pm$0.05 & 8.679$\pm$0.133 & 8.099$\pm$0.045 &6.039&5.450&4.815&4.214&3.983&3.881& 3.853 & 3.849 & 6.44\\
QZ Nor \tablenotemark{\scriptsize a} &0.578&0.267& 0.484$\pm$0.020 & 0.79$\pm$0.06 & 11.586$\pm$0.081 & 10.512$\pm$0.165 &9.752&8.859&7.858&7.080&6.749&6.598& \nodata & \nodata & 1.03\\
SU Cyg \tablenotemark{\scriptsize a}\tablenotemark{\scriptsize b} &0.585&0.103& 1.055$\pm$0.052 & 1.19$\pm$0.23 & 9.872$\pm$0.101 & 9.622$\pm$0.425 &7.427&6.859&6.184&5.634&5.399&5.300& \nodata & \nodata & 3.44\\
Y Lac \tablenotemark{\scriptsize a} &0.636&0.218& 0.431$\pm$0.013 & 0.44$\pm$0.02 & 11.811$\pm$0.056 & 11.783$\pm$0.099 &9.868&9.141&8.288&7.647&7.323&7.201& \nodata & \nodata & 1.05\\
T Vul \tablenotemark{\scriptsize a} &0.647&0.068& 1.719$\pm$0.058 & 1.90$\pm$0.02 & 8.826$\pm$0.066 & 8.606$\pm$0.023 &6.397&5.753&5.072&4.546&4.283&4.181& 4.114 & 4.111 & 1.20\\
FF Aql \tablenotemark{\scriptsize a} &0.650&0.207& 1.938$\pm$0.071 & 2.81$\pm$0.03 & 8.584$\pm$0.074 & 7.756$\pm$0.023 &6.126&5.370&4.501&3.851&3.580&3.461& 3.379 & 3.353 & 1.06\\
T Vel \tablenotemark{\scriptsize a} &0.667&0.305& 0.940$\pm$0.016 & 0.99$\pm$0.02 & 10.141$\pm$0.033 & 10.022$\pm$0.044 &8.964&8.025&6.948&6.163&5.779&5.622& \nodata & \nodata & 0.93\\
VZ Cyg \tablenotemark{\scriptsize a} &0.687&0.281& 0.545$\pm$0.016 & 0.54$\pm$0.01 & 11.304$\pm$0.069 & 11.338$\pm$0.040 &9.843&8.967&7.979&7.228&6.892&6.739& \nodata & \nodata & 1.31\\
V350 Sgr \tablenotemark{\scriptsize b} &0.712&0.315& 0.810$\pm$0.062 & 1.07$\pm$0.01 & 10.496$\pm$0.161 & 9.853$\pm$0.020 &8.378&7.479&6.419&5.624&5.247&5.117& \nodata & \nodata & 2.43\\
BG Lac &0.727&0.316& 0.581$\pm$0.019 & 0.59$\pm$0.05 & 11.176$\pm$0.078 & 11.146$\pm$0.184 &9.855&8.895&7.824&7.068&6.680&6.530& \nodata & \nodata & 1.43\\
$\delta$ Cep \tablenotemark{\scriptsize b} &0.730&0.079& 3.578$\pm$0.148 & 3.66$\pm$0.01 & 7.234$\pm$0.076 & 7.183$\pm$0.006 &4.620&3.955&3.220&2.703&2.406&2.301& 2.221 & 2.217 & 2.71\\
CV Mon &0.731&0.762& 0.601$\pm$0.015 & 0.63$\pm$0.15 & 11.109$\pm$0.064 & 11.003$\pm$0.527 &11.597&10.295&8.638&7.332&6.802&6.558& 6.375 & 6.360 & 1.10\\
V Cen &0.740&0.308& 1.409$\pm$0.022 & 1.65$\pm$0.05 & 9.254$\pm$0.032 & 8.913$\pm$0.066 &7.698&6.826&5.805&5.027&4.652&4.504& 4.405 & 4.400 & 1.06\\
Y Sgr &0.761&0.202& 2.012$\pm$0.058 & 2.13$\pm$0.03 & 8.487$\pm$0.065 & 8.358$\pm$0.031 &6.596&5.745&4.782&4.088&3.719&3.601& 3.486 & 3.483 & 1.76\\
CS Vel &0.771&0.778& 0.272$\pm$0.012 & 0.33$\pm$0.18 & 12.824$\pm$0.102 & 12.407$\pm$1.329 &13.045&11.698&10.062&8.770&8.245&7.997& \nodata & \nodata & 0.91\\
BB Sgr &0.822&0.296& 1.188$\pm$0.024 & 1.27$\pm$0.19 & 9.626$\pm$0.037 & 9.481$\pm$0.327 &7.930&6.939&5.842&5.045&4.654&4.508& \nodata & \nodata & 0.82\\
V Car &0.826&0.178& 0.797$\pm$0.014 & 0.96$\pm$0.04 & 10.487$\pm$0.043 & 10.089$\pm$0.091 &8.224&7.345&6.410&5.748&5.397&5.263& \nodata & \nodata & 1.04\\
U Sgr &0.829&0.425& 1.605$\pm$0.022 & 1.77$\pm$0.20 & 8.973$\pm$0.030 & 8.760$\pm$0.246 &7.795&6.702&5.450&4.529&4.104&3.943& 3.824 & 3.822 & 0.85\\
V496 Aql &0.833&0.419& 0.977$\pm$0.034 & 1.07$\pm$0.18 & 10.029$\pm$0.078 & 9.853$\pm$0.369 &8.902&7.745&6.471&5.556&5.133&4.984& \nodata & \nodata & 1.56\\
X Sgr &0.846&0.250& 2.843$\pm$0.140 & 3.00$\pm$0.06 & 7.745$\pm$0.090 & 7.614$\pm$0.043 &5.307&4.560&3.649&2.967&2.652&2.534& 2.423 & 2.409 & 1.22\\
U Aql \tablenotemark{\scriptsize b} &0.847&0.380& 1.765$\pm$0.087 & 1.46$\pm$0.05 & 8.758$\pm$0.088 & 9.178$\pm$0.074 &7.465&6.425&5.268&4.381&4.001&3.839& 3.738 & 3.736 & 3.09\\
$\eta$ Aql \tablenotemark{\scriptsize b} &0.856&0.137& 3.711$\pm$0.194 & 4.15$\pm$0.06 & 7.170$\pm$0.106 & 6.910$\pm$0.031 &4.690&3.900&3.025&2.402&2.075&1.959& 1.864 & 1.865 & 2.56\\
W Sgr \tablenotemark{\scriptsize b} &0.881&0.114& 2.402$\pm$0.177 & 2.28$\pm$0.02 & 8.162$\pm$0.144 & 8.210$\pm$0.019 &5.417&4.669&3.842&3.212&2.893&2.781& 2.721 & 2.719 & 3.95\\
U Vul \tablenotemark{\scriptsize b} &0.903&0.636& 1.308$\pm$0.057 & 1.46$\pm$0.29 & 9.424$\pm$0.091 & 9.178$\pm$0.437 &8.409&7.136&5.610&4.575&4.118&3.947& 3.797 & 3.778 & 2.88\\
S Sge \tablenotemark{\scriptsize b} &0.923&0.105& 1.700$\pm$0.111 & 1.48$\pm$0.13 & 8.889$\pm$0.112 & 9.149$\pm$0.191 &6.412&5.612&4.775&4.173&3.857&3.747& 3.652 & 3.661 & 4.00\\
GH Lup &0.967&0.353& 0.864$\pm$0.021 & 0.89$\pm$0.11 & 10.314$\pm$0.063 & 10.253$\pm$0.270 &8.840&7.625&6.350&5.428&4.977&4.790& \nodata & \nodata & 0.95\\
S Mus \tablenotemark{\scriptsize b} &0.985&0.224& 1.179$\pm$0.092 & 1.22$\pm$0.08 & 9.670$\pm$0.177 & 9.568$\pm$0.143 &6.965&6.123&5.184&4.497&4.141&3.989& \nodata & \nodata & 4.49\\
S Nor &0.989&0.189& 1.099$\pm$0.022 & 1.06$\pm$0.24 & 9.801$\pm$0.046 & 9.873$\pm$0.500 &7.381&6.432&5.424&4.682&4.297&4.151& 4.066 & 4.085 & 0.88\\
$\beta$ Dor \tablenotemark{\scriptsize b} &0.993&0.055& 2.937$\pm$0.139 & 3.14$\pm$0.05 & 7.666$\pm$0.100 & 7.515$\pm$0.035 &4.555&3.753&2.937&2.394&2.056&1.945& 1.858 & 1.871 & 4.53\\
$\zeta$ Gem \tablenotemark{\scriptsize b} &1.007&0.015& 3.112$\pm$0.218 & 2.78$\pm$0.03 & 7.550$\pm$0.198 & 7.780$\pm$0.023 &4.709&3.895&3.107&2.483&2.178&2.075& 2.025 & 2.037 & 2.78\\
Z Lac &1.037&0.390& 0.510$\pm$0.021 & 0.53$\pm$0.01 & 11.475$\pm$0.070 & 11.379$\pm$0.041 &9.534&8.434&7.214&6.306&5.863&5.689& \nodata & \nodata & 1.05\\
XX Cen &1.040&0.281& 0.570$\pm$0.026 & 0.66$\pm$0.18 & 11.240$\pm$0.089 & 10.902$\pm$0.608 &8.807&7.831&6.744&5.952&5.557&5.403& \nodata & \nodata & 1.24\\
V340 Nor &1.053&0.339& 0.491$\pm$0.025 & 0.56$\pm$0.16 & 11.566$\pm$0.103 & 11.259$\pm$0.638 &9.517&8.356&7.158&6.201&5.745&5.561& 5.453 & 5.480 & 0.92\\
UU Mus &1.066&0.421& 0.306$\pm$0.012 & 0.33$\pm$0.20 & 12.560$\pm$0.076 & 12.407$\pm$1.526 &10.955&9.806&8.509&7.492&7.039&6.839& \nodata & \nodata & 1.01\\
U Nor &1.102&0.909& 0.625$\pm$0.019 & 0.81$\pm$0.14 & 11.019$\pm$0.060 & 10.458$\pm$0.379 &10.844&9.232&7.347&5.868&5.258&4.985& \nodata & \nodata & 0.98\\
SU Cru \tablenotemark{\scriptsize d} &1.109&0.994& 0.178$\pm$0.145 & 0.62$\pm$0.03 & 14.032$\pm$0.786 & 11.038$\pm$0.105 &11.553&9.781&7.654&5.934&5.054&4.736& \nodata & \nodata & 1.52\\
BN Pup &1.136&0.439& 0.301$\pm$0.015 & 0.26$\pm$0.01 & 12.645$\pm$0.118 & 12.925$\pm$0.084 &11.045&9.849&8.510&7.526&7.056&6.863& \nodata & \nodata & 1.25\\
TT Aql &1.138&0.462& 0.998$\pm$0.022 & 0.99$\pm$0.04 & 10.012$\pm$0.046 & 10.022$\pm$0.088 &8.445&7.137&5.730&4.714&4.226&4.038& 3.875 & 3.909 & 1.08\\
LS Pup &1.151&0.486& 0.214$\pm$0.016 & 0.21$\pm$0.01 & 13.407$\pm$0.176 & 13.389$\pm$0.103 &11.709&10.478&9.092&8.062&7.566&7.374& \nodata & \nodata & 1.25\\
VW Cen &1.177&0.452& 0.260$\pm$0.016 & 0.28$\pm$0.03 & 12.936$\pm$0.143 & 12.764$\pm$0.234 &11.635&10.277&8.773&7.617&7.062&6.827& \nodata & \nodata & 1.06\\
X Cyg &1.214&0.241& 0.910$\pm$0.020 & 0.86$\pm$0.03 & 10.206$\pm$0.046 & 10.328$\pm$0.076 &7.532&6.404&5.244&4.402&3.980&3.814& 3.678 & 3.728 & 1.28\\
CD Cyg &1.232&0.520& 0.394$\pm$0.016 & 0.39$\pm$0.03 & 12.027$\pm$0.083 & 12.045$\pm$0.167 &10.260&8.957&7.511&6.378&5.881&5.682& 5.477 & 5.530 & 1.01\\
SZ Aql &1.234&0.567& 0.525$\pm$0.020 & 0.46$\pm$0.03 & 11.404$\pm$0.085 & 11.686$\pm$0.142 &10.058&8.631&7.059&5.891&5.368&5.150& 4.981 & 5.032 & 0.94\\
Y Oph \tablenotemark{\scriptsize c} &1.234&0.680& 1.348$\pm$0.036 & 1.81$\pm$0.01 & 9.379$\pm$0.053 & 8.712$\pm$0.012 &7.550&6.175&4.546&3.358&2.876&2.677& 2.528 & 2.500 & 1.03\\
VY Car &1.277&0.250& 0.565$\pm$0.017 & 0.55$\pm$0.03 & 11.234$\pm$0.059 & 11.298$\pm$0.119 &8.611&7.454&6.253&5.391&4.946&4.778& \nodata & \nodata & 0.92\\
RU Sct &1.294&0.972& 0.526$\pm$0.024 & 0.58$\pm$0.02 & 11.409$\pm$0.094 & 11.183$\pm$0.075 &11.139&9.468&7.478&5.959&5.337&5.073& 4.856 & 4.873 & 0.87\\
RY Sco &1.308&0.757& 0.764$\pm$0.032 & 0.85$\pm$0.02 & 10.579$\pm$0.075 & 10.353$\pm$0.051 &9.480&8.022&6.280&4.938&4.389&4.136& \nodata & \nodata & 0.73\\
RZ Vel &1.310&0.315& 0.661$\pm$0.017 & 0.70$\pm$0.02 & 10.913$\pm$0.049 & 10.775$\pm$0.062 &8.209&7.080&5.851&4.926&4.478&4.298& \nodata & \nodata & 1.24\\
WZ Sgr &1.339&0.455& 0.612$\pm$0.028 & 0.57$\pm$0.05 & 11.097$\pm$0.105 & 11.221$\pm$0.191 &9.414&8.013&6.515&5.326&4.771&4.544& 4.364 & 4.443 & 0.94\\
WZ Car &1.362&0.390& 0.284$\pm$0.018 & 0.29$\pm$0.08 & 12.752$\pm$0.156 & 12.688$\pm$0.615 &10.423&9.273&7.983&6.956&6.483&6.289& \nodata & \nodata & 1.38\\
SW Vel &1.370&0.363& 0.413$\pm$0.018 & 0.42$\pm$0.02 & 11.930$\pm$0.090 & 11.884$\pm$0.103 &9.269&8.115&6.828&5.867&5.403&5.214& \nodata & \nodata & 1.05\\
T Mon &1.432&0.191& 0.745$\pm$0.052 & 0.72$\pm$0.01 & 10.702$\pm$0.151 & 10.713$\pm$0.030 &7.299&6.128&4.988&4.133&3.678&3.512& 3.359 & 3.425 & 1.72\\
RY Vel &1.449&0.577& 0.376$\pm$0.021 & 0.45$\pm$0.04 & 12.162$\pm$0.120 & 11.734$\pm$0.194 &9.728&8.361&6.816&5.628&5.131&4.902& \nodata & \nodata & 1.08\\
AQ Pup &1.479&0.546& 0.294$\pm$0.023 & 0.33$\pm$0.04 & 12.674$\pm$0.174 & 12.407$\pm$0.265 &10.039&8.691&7.143&6.044&5.508&5.294& \nodata & \nodata & 1.18\\
KN Cen &1.532&0.841& 0.251$\pm$0.018 & 0.27$\pm$0.15 & 13.009$\pm$0.163 & 12.843$\pm$1.360 &11.424&9.827&7.975&6.442&5.775&5.476& \nodata & \nodata & 1.03\\
l Car \tablenotemark{\scriptsize b} &1.551&0.155& 1.988$\pm$0.110 & 2.01$\pm$0.12  & 8.537$\pm$0.135 & 8.484$\pm$0.130 &4.996&3.739&2.562&1.712&1.239&1.080& 0.925 & 1.047 & 2.39\\
U Car &1.589&0.280& 0.561$\pm$0.023 & 0.67$\pm$0.10 & 11.262$\pm$0.069 & 10.870$\pm$0.327 &7.481&6.296&5.069&4.149&3.703&3.520& 3.357 & 3.415 & 1.23\\
RS Pup &1.617&0.482& 0.581$\pm$0.017 & 0.55$\pm$0.11 & 11.190$\pm$0.063 & 11.298$\pm$0.440 &8.451&7.014&5.474&4.358&3.823&3.602& \nodata & \nodata & 1.16\\
SV Vul \tablenotemark{\scriptsize e} &1.653&0.486& 0.402$\pm$0.021 & 0.46$\pm$0.08 & 12.009$\pm$0.114 & 11.686$\pm$0.382 &8.666&7.205&5.690&4.600&4.094&3.901& 3.711 & 3.788 & 1.20\\
\enddata

\tablenotetext{\scriptsize a}{Removed: $\log P < 0.7$}
\tablenotetext{\scriptsize b}{Removed: {\tt ruwe} $> 2.0$}
\tablenotetext{\scriptsize c}{Removed: classified as an overtone pulsator in the \textit{General Catalog of Variable Stars}}
\tablenotetext{\scriptsize d}{Removed: large fractional parallax uncertainty ($\sigma_{\pi} / \pi$) in EDR3 Catalog}
\tablenotetext{\scriptsize e}{Removed: significant outlier in PL relations}
\tablerefs{\cite{2020arXiv201205220B, 2002AJ....124.1695B, 2007AJ....133.1810B, 2007AA...476...73F, 2012ApJ...759..146M}}

\end{deluxetable*}
\end{longrotatetable}

The sample described above is comprised of the brightest, nearest, and lowest extinction Cepheids in the Milky Way. All of these Cepheids have dozens of phase points available in the optical and near-infrared from observations over several decades, and the small amplitudes and photometric errors in the mid-IR give exquisitely accurate mean magnitudes. All Cepheids also have radial velocity measurements, which were used for determining individual Baade-Wesselink distances in \cite{2007AA...476...73F}.

\subsection{Large Magellanic Cloud}\label{subsec:LMCphot}

In the LMC, we analyzed $BVI_{c}$ mean magnitudes from the Cepheid sample of the OGLE~II survey \citep{1999AcA....49..201U}. We chose the OGLE~II data over the newer OGLE~III/OGLE~IV sample given the availability of homogeneous $B$-band data in OGLE~II. However, we also provide the results using $VI_{c}$ data from OGLE~III \citep{2008AcA....58..163S}. We also added $BVI_{c}$ data for the 66 unique Cepheids from \cite{2002ApJS..142...71S} to both samples. The Sebo et al. sample has the advantage of including 8 additional Cepheids with periods longer than 30 days.

For the near-infrared, we used the $JHK_{s}$ magnitudes from \cite{2004AJ....128.2239P}. They observed 92 Cepheids with an average of 22 phase points with the Swope and du~Pont telescopes at the Las Campanas Observatory, fitting the light curves using a locally-weighted-regression smoother, and using periods adopted from the literature. Our $[3.6]$ and $[4.5]$ micron data were taken over 24 epochs with warm \textit{Spitzer} as published by \cite{2011ApJ...743...76S}. 

\subsection{Small Magellanic Cloud}\label{subsec:SMCphot}

We used data in the SMC from \cite{2016ApJ...816...49S} for all wavelengths. They compiled light curve data for $B$ through $K$ from several sources in the literature (see their Table 3) and calculated mean magnitudes using a Gaussian local estimation algorithm. They also collected warm \textit{Spitzer} data in the $[3.6]$ and $[4.5]$ micron bands at 12 epochs for a sample of 90 fundamental-mode Cepheids.

\subsection{Photometric Sample Refinement}\label{subsec:refphot}

In all three cases we limited the samples to Cepheids with periods longer than 5 days ($\log P > 0.7$). This particularly affects the $BVI_{c}$ samples from OGLE in the SMC and LMC, as they cataloged large numbers of short-period Cepheids which lie well below our period cut-off. This helps to mitigate possible non-linearities in the PL relation since our Galactic Cepheids more uniformly sample the chosen period range.

The Milky Way Cepheid sample was further constrained by removing stars with the goodness-of-fit parameter {\tt ruwe} $>2.0$, following the less-stringent suggested cut of \cite{2021arXiv210110206M}. We also removed a suspected overtone pulsator (Y~Oph), one star with an extremely high fractional parallax error (SU~Cru), and a PL relation outlier (SV~Vul). This leaves a final sample of 37 Cepheids in the visible and near-IR and 14 Cepheids in the mid-IR. For more specific details on sample selection, we refer the reader to Appendix~\ref{sec:A_refinement}.

We note that the optical and near-infrared samples of Cepheids in the LMC do not have good overlap; there are only 20 Cepheids in common, and most of these have periods shorter than the period cutoff. Moreover, the $[3.6]$ and $[4.5]$ micron bands comprise only a subset of the $JHK_{s}$ Cepheids. Thus, since the overlapping sample does not contain enough long-period Cepheids to constrain the PL relation, we instead separately consider the optical and IR samples. In principle, this could introduce systematic effects, particularly due to the difference in period distributions. For this reason, we also provide the results of fits to the homogeneous set of $JHK_{s}$ and $[3.6]$ photometry, for comparison. As described in Section \ref{subsec:EDR3dist}, restricting to the homogeneous sample affects the LMC distance moduli by $<0.02$ mag, so using the heterogeneous sample does not seem to introduce significant systematic errors.

\section{EDR3 Distance Determinations}\label{sec:MWdist}

In Section \ref{subsec:Gaia_MW_dists}, we discuss obtaining distances to the individual Milky Way Cepheids from \textit{Gaia} EDR3, adopting published reddenings from the literature. In Section \ref{subsec:EDR3dist}, we describe obtaining the distance moduli and reddenings to the LMC and SMC using a multi-wavelength reddening law fit; then we also find the distance moduli using multiple formulations of the reddening-free Wesenheit function in Section \ref{sec:Wes_DMs}.

\subsection{Individual Milky Way Distances}\label{subsec:Gaia_MW_dists}

All stars in the Milky Way sample were first cross-matched with the \textit{Gaia} EDR3 catalog, using a search radius of 5~arcsec and eliminating extraneous stars along the line-of-sight based on their parallaxes and apparent \textit{G} magnitudes.

For each Milky Way Cepheid, we used the unique \textit{Gaia} source ID to obtain their photogeometric distance estimates from \cite{2020arXiv201205220B}. This distance measure is calculated using the ($BP-RP$) colors, $G$ magnitudes, and EDR3 parallaxes, applying two direction-dependent priors. The ``geometric" prior accounts for the distribution of stellar distances along a line of sight, while the ``photometric" prior accounts for the distribution of absolute extincted stellar magnitudes along that line of sight. 

We adopt a standard reddening law from \cite{1989ApJ...345..245C} with a reddening coefficient $R_{V} = 3.1$ for our $BVI_{c}$ and $JHK_{s}$ magnitudes. For the $[3.6]$ and $[4.5]$ micron bands, we use the reddening law from \cite{2005ApJ...619..931I}, which is calibrated using field stars in the Galactic plane using \textit{Spitzer} and 2MASS data. This is consistent with the analysis of \cite{2012ApJ...759..146M}, from which we obtained the mid-IR data. 

\begin{figure}
    \centering
    \includegraphics[width=\columnwidth]{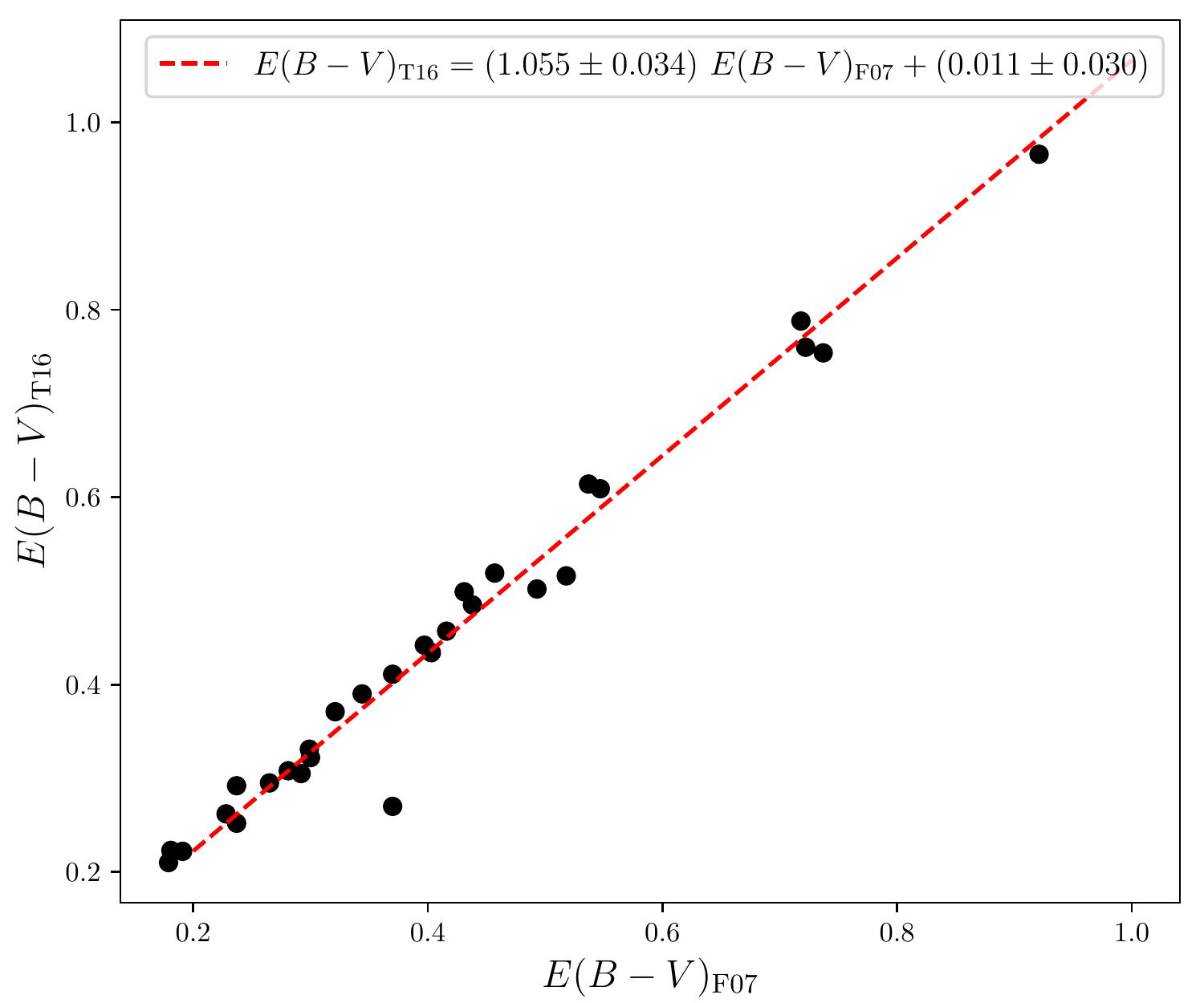}
    \caption{Comparison between the reddenings of \cite{2007AA...476...73F} (F07) and \cite{2016RMxAA..52..223T} (T16). We find a significant scaling term, which we apply to all of the F07 values to obtain corrected reddenings.}
    \label{fig:EBV_comp}
\end{figure}

Individual extinctions were obtained from \cite{2007AA...476...73F}. They gathered extinctions from the database of \cite{1995IBVS.4148....1F}, converting them to the \cite{2007MNRAS.377..147L} system, and then taking an error-weighted mean. We adjusted these reddenings to the more recent system of \cite{2016RMxAA..52..223T} based on an overlapping sample of 29 stars.  We find a significant scaling factor of $1.055\pm0.034$ must be applied to the \citealt{2007AA...476...73F} reddenings, as shown in Figure \ref{fig:EBV_comp}. However, we choose not to apply the small zero-point offset of $+0.011 \pm 0.030$, as it is statistically consistent with zero. The systematic errors due to this conversion are calculated in Section \ref{subsubsec:red_adj}.

\begin{figure*}
    \centering
    \includegraphics[width=\textwidth]{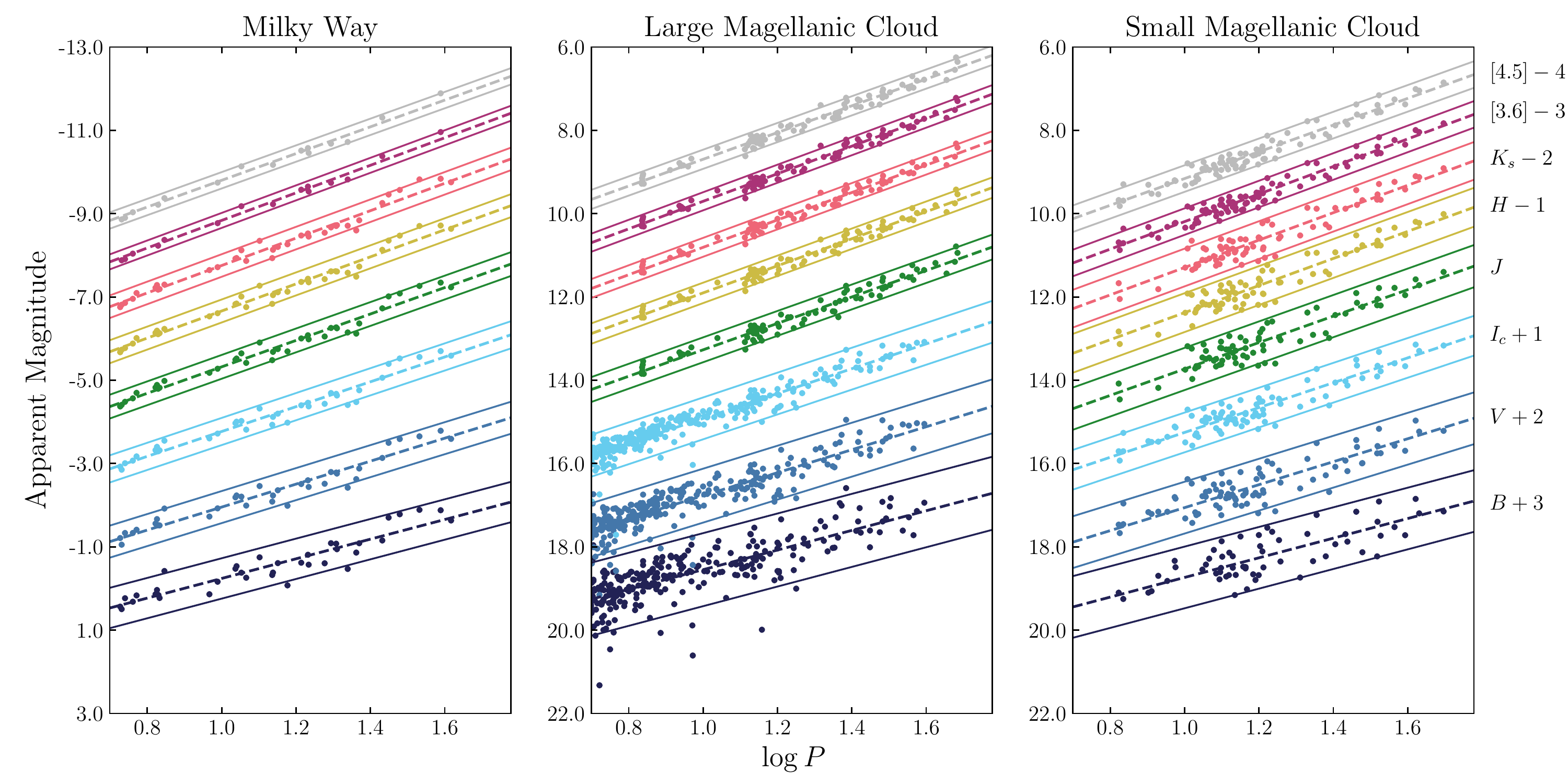}
    \caption{Period-Luminosity relations for Cepheids in the Milky Way, LMC, and SMC in the $BVI_{c}$, $JHK_{s}$, and $[3.6]$ \& $[4.5]$ micron bands. Dashed lines show the ridge of the PL relations, and solid lines show the 2$\sigma$ uncertainties. The Milky Way Cepheids' absolute magnitudes have been corrected for individual line-of-sight reddenings. Fixed slopes from the LMC were fit to each band in the period range from 5 to 60 days. Individual bands are shifted vertically for visual distinction, by the amounts noted in the far right-hand labels.\label{fig:PL_rels}}
\end{figure*}

\begin{deluxetable*}{lccccccccc}
\label{tab:final_PLRs}
\tablecaption{Adopted Fixed-Slope Period-Luminosity Relations and LMC/SMC Apparent Distance Moduli}

\tablehead{\colhead{Band} & \colhead{Fixed Slope} & \multicolumn{2}{c}{Milky Way} & \multicolumn{3}{c}{LMC} & \multicolumn{3}{c}{SMC} \\ \cmidrule(lr){3-4} \cmidrule(lr){5-7} \cmidrule(lr){8-10}
\colhead{} & \colhead{a} & \colhead{b} & \colhead{$\sigma$} & \colhead{b} & \colhead{$\sigma$} & \colhead{$\mu$} & \colhead{b} & \colhead{$\sigma$} & \colhead{$\mu$}} 

\startdata
$B$ & $-2.365\pm0.119$ & $-3.736\pm0.042$ & 0.258 & $15.067\pm0.027$ & 0.445 & $18.803\pm0.050$ & $15.263\pm0.048$ & 0.370 & $18.998\pm0.064$ \\
$V$ & $-2.762\pm0.088$ & $-4.505\pm0.034$ & 0.205 & $14.212\pm0.020$ & 0.330 & $18.716\pm0.039$ & $14.507\pm0.034$ & 0.311 & $19.011\pm0.048$ \\
$I_{c}$ & $-2.987\pm0.068$ & $-5.337\pm0.029$ & 0.176 & $13.313\pm0.015$ & 0.257 & $18.650\pm0.033$ & $13.657\pm0.026$ & 0.238 & $18.994\pm0.039$ \\
$J$ & $-3.144\pm0.074$ & $-5.901\pm0.025$ & 0.154 & $12.637\pm0.017$ & 0.149 & $18.357\pm0.030$ & $13.096\pm0.030$ & 0.253 & $18.997\pm0.039$ \\
$H$ & $-3.224\pm0.061$ & $-6.251\pm0.024$ & 0.146 & $12.254\pm0.014$ & 0.123 & $18.505\pm0.028$ & $12.728\pm0.027$ & 0.233 & $18.979\pm0.036$ \\
$K_{s}$ & $-3.265\pm0.056$ & $-6.346\pm0.023$ & 0.142 & $12.152\pm0.013$ & 0.113 & $18.498\pm0.027$ & $12.637\pm0.027$ & 0.226 & $18.983\pm0.035$ \\
$[3.6]$ & $-3.284\pm0.054$ & $-6.450\pm0.022$ & 0.087 & $12.043\pm0.012$ & 0.109 & $18.494\pm0.026$ & $12.531\pm0.017$ & 0.159 & $18.981\pm0.028$ \\
$[4.5]$ & $-3.182\pm0.057$ & $-6.393\pm0.025$ & 0.097 & $12.054\pm0.013$ & 0.115 & $18.447\pm0.028$ & $12.517\pm0.017$ & 0.160 & $18.910\pm0.031$ \\
\enddata

\tablecomments{The form of the PL relation is $M_{\lambda} = a (\log P - 1.2) + b$, where $a$ is the fixed slope from the LMC. The apparent distance moduli ($\mu$) are found by subtracting the LMC and SMC intercepts from the Milky Way calibration. Milky Way values and calculated distance moduli include a +18 $\mu$as offset.}

\end{deluxetable*}

\subsection{Reddening Curve Fit Distance Moduli}\label{subsec:EDR3dist}

\begin{deluxetable}{c|lccc}
\tablecaption{Unfixed-Slope Period-Luminosity Relations\label{tab:unfixed_slope}}

\tablehead{\colhead{} & \colhead{Filter} & \colhead{a} & \colhead{b} & \colhead{$\sigma$}} 

\startdata
\multirow{8}{*}{Milky Way} & $B$ & $-2.341\pm0.168$ & $-3.734\pm0.042$ & 0.258 \\
 & $V$ & $-2.656\pm0.133$ & $-4.497\pm0.033$ & 0.204 \\
 & $I_{c}$ & $-2.917\pm0.115$ & $-5.332\pm0.029$ & 0.176 \\
 & $J$ & $-3.121\pm0.100$ & $-5.899\pm0.025$ & 0.153 \\
 & $H$ & $-3.240\pm0.095$ & $-6.252\pm0.024$ & 0.146 \\
 & $K_{s}$ & $-3.285\pm0.093$ & $-6.347\pm0.023$ & 0.142 \\
 & $[3.6]$ & $-3.372\pm0.090$ & $-6.460\pm0.022$ & 0.085 \\
 & $[4.5]$ & $-3.268\pm0.102$ & $-6.402\pm0.025$ & 0.096 \\
 \hline
\multirow{8}{*}{SMC} & $B$ & $-2.184\pm0.234$ & $15.260\pm0.048$ & 0.368 \\
 & $V$ & $-2.520\pm0.170$ & $14.509\pm0.034$ & 0.307 \\
 & $I_{c}$ & $-2.873\pm0.133$ & $13.658\pm0.026$ & 0.237 \\
 & $J$ & $-3.155\pm0.159$ & $13.095\pm0.030$ & 0.253 \\
 & $H$ & $-3.344\pm0.146$ & $12.728\pm0.027$ & 0.233 \\
 & $K_{s}$ & $-3.340\pm0.142$ & $12.636\pm0.026$ & 0.226 \\
 & $[3.6]$ & $-3.299\pm0.089$ & $12.530\pm0.017$ & 0.159 \\
 & $[4.5]$ & $-3.239\pm0.089$ & $12.516\pm0.017$ & 0.159 \\
\hline
\multirow{2}{*}{LMC\tablenotemark{\scriptsize a}} & $V$ & $-2.779\pm0.063$ & $14.220\pm0.014$ & 0.334 \\
 & $I_{c}$ & $-2.942\pm0.055$ & $13.333\pm0.011$ & 0.252 \\
\enddata

\tablecomments{The form of the PL relation is $M_{\lambda} = a (\log P - 1.2)~+~b$. Milky Way values include a parallax offset of +18~$\mu$as. The adopted LMC unfixed-slope fits are found in Columns 2, 5 and 6 of Table~2.}

\tablenotetext{\scriptsize a}{These values are from the OGLE~III sample of Cepheids, and are provided for comparison with the adopted values in Table \ref{tab:final_PLRs} from the OGLE~II sample of Cepheids.} 
\end{deluxetable}

\begin{figure}
    \centering
    \includegraphics[width=\columnwidth]{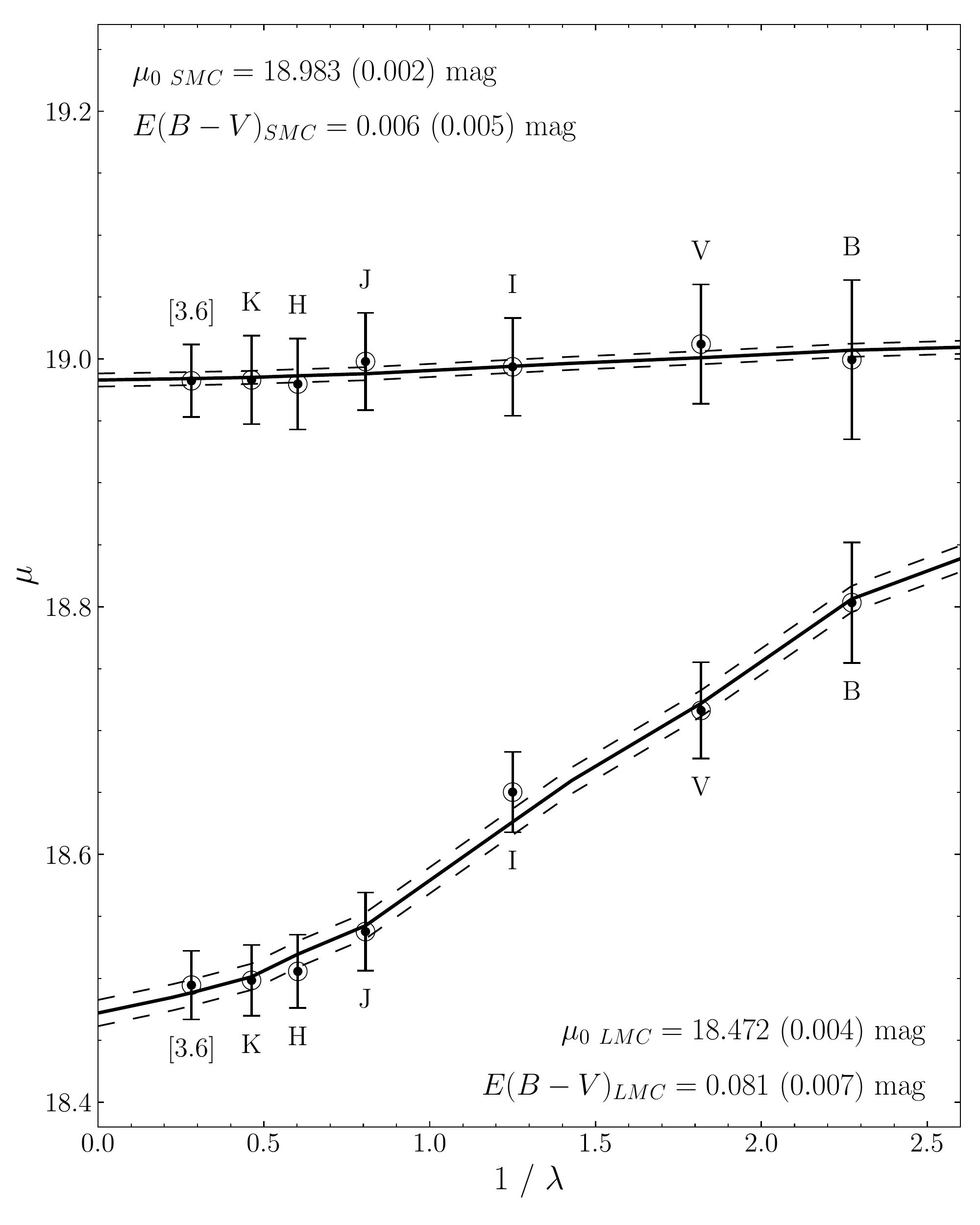}
    \caption{Reddening and distance modulus fit for the SMC (top) and LMC (bottom) based on the Milky Way Cepheid PL relation calibration. The dashed lines show the 2$\sigma$ scatter on each fit, and the errors are the standard error of the mean. Values include a +18.0 $\mu$as offset. The [4.5] micron band is excluded from the fit due to the metallicity-sensitive CO bandhead in that wavelength range.}
    \label{fig:final_DMs}
\end{figure}

We show PL relations for all three galaxies in Figure \ref{fig:PL_rels}. The fixed-slope fits using OGLE~II data for the optical sample in the LMC are given in Table \ref{tab:final_PLRs}. We additionally provide the fits with unfixed-slopes, including the OGLE III sample of optical Cepheids in the LMC in Table \ref{tab:unfixed_slope}. We use the absolute calibration of the Milky Way period-luminosity relations to determine distances to the Small and Large Magellanic Clouds. We subtract the Milky Way intercepts at $\log P = 1.2$ from the respective intercepts of the SMC and LMC, giving an estimation of the apparent distance modulus for each band, as given in Table \ref{tab:final_PLRs}. These wavelength-dependent apparent distance moduli are then fit with our adopted reddening curve.

Often in the literature, the PL relation is fit about a pivot point of $\log P = 0.0$ or $\log P = 1.0$. Fitting about 0.0 has the advantage of mathematical simplicity, and for fixed slope PL relations, the choice of pivot has no effect on the derived distances. However, when performing unfixed-slope PL fits, shifting the fit-axis closer to the median period decouples the error in the calculated value of the intercept from the value of the slope. This can be important when comparing samples with different apparent slopes, which is often the case in extragalactic Cepheid samples. We choose to pivot at $\log P = 1.2$ because it is the mid-point of our adopted period range ($0.7 < \log P < 1.7$), most effectively minimizing the aforementioned coupling. We also note that changing the value of the pivot to either $\log P = 0.7$ or $1.0$ (as in i.e. \citealt{2021arXiv210310894B, 2021ApJ...908L...6R}) has an effect of $<0.02$ mag on all unfixed-slope distance moduli.

While we have data in the $[4.5]$ micron band for all three galaxies, $\mu_{[4.5]}$ is not included in the reddening law fits. This wavelength is known to be contaminated by a rotational-vibrational CO bandhead \citep{2010ApJ...709..120M, 2011ApJ...743...76S}. The depth of this bandhead correlates with metallicity \citep{2016MNRAS.459.1170S}, and since our three galaxies vary significantly in metallicity, the $[4.5]$ band is poorly suited for distance determinations. The PL fits are included for completeness, but not used thereafter.

\begin{deluxetable*}{cccccccccccrr}
\tablecaption{Distance Moduli and Extinctions from Reddening Law Fits to Samples in the LMC \label{tab:LMC_final}}

\tablehead{\colhead{Slopes} & \colhead{OGLE} & \colhead{Offset} & \colhead{$\mu_{B}$} & \colhead{$\mu_{V}$} & \colhead{$\mu_{I}$} & \colhead{$\mu_{J}$} & \colhead{$\mu_{H}$} & \colhead{$\mu_{K}$} & \colhead{$\mu_{[3.6]}$} & \colhead{$\mu_{[4.5]}$} & \colhead{$\mu_{0}$} & \colhead{$E(B-V)$} \\ 
\colhead{} & \colhead{} & \colhead{($\mu$as)} & \colhead{(mag)} & \colhead{(mag)} & \colhead{(mag)} & \colhead{(mag)} & \colhead{(mag)} & \colhead{(mag)} & \colhead{(mag)} & \colhead{(mag)} & \colhead{(mag)} & \colhead{(mag)}} 

\startdata
fixed LMC & II & 0 & $18.876$ & $18.789$ & $18.724$ & $18.611$ & $18.579$ & $18.571$ & $18.544$ & $18.497$ & $18.537\pm0.004$ & $0.083\pm0.004$ \\
 & (BVI excl.) &  &  &  &  &  &  &  &  &  & $18.530\pm0.003$ & $0.092\pm0.015$ \\
 &  & $+18$ & $18.803$ & $18.716$ & $18.650$ & $18.537$ & $18.505$ & $18.498$ & $18.494$ & $18.447$ & $18.472\pm0.004$ & $0.081\pm0.004$ \\
 & (BVI excl.) &  &  &  &  &  &  &  &  &  & $18.476\pm0.003$ & $0.064\pm0.015$ \\
 & III & 0 & \nodata & $18.798$ & $18.740$ & $18.611$ & $18.579$ & $18.571$ & $18.544$ & $18.497$ & $18.534\pm0.006$ & $0.091\pm0.008$ \\
 &  & $+18$ & \nodata & $18.725$ & $18.666$ & $18.537$ & $18.505$ & $18.498$ & $18.494$ & $18.447$ & $18.468\pm0.006$ & $0.088\pm0.008$ \\
\hline
unfixed & II & 0 & $18.879$ & $18.786$ & $18.723$ & $18.613$ & $18.584$ & $18.577$ & $18.559$ & $18.512$ & $18.545\pm0.004$ & $0.081\pm0.003$ \\
 & (BVI excl.) &  &  &  &  &  &  &  &  &  & $18.544\pm0.002$ & $0.076\pm0.011$ \\
 &  & $+18$ & $18.802$ & $18.709$ & $18.645$ & $18.536$ & $18.506$ & $18.500$ & $18.503$ & $18.456$ & $18.474\pm0.004$ & $0.079\pm0.004$ \\
 & (BVI excl.) &  &  &  &  &  &  &  &  &  & $18.486\pm0.003$ & $0.051\pm0.019$ \\
 & III & 0 & \nodata & $18.794$ & $18.742$ & $18.613$ & $18.584$ & $18.577$ & $18.559$ & $18.512$ & $18.544\pm0.007$ & $0.086\pm0.008$ \\
 &  & $+18$ & \nodata & $18.716$ & $18.665$ & $18.536$ & $18.506$ & $18.500$ & $18.503$ & $18.456$ & $18.474\pm0.007$ & $0.083\pm0.008$ \\
\hline
fixed MW & II & 0 & $18.870$ & $18.797$ & $18.725$ & $18.615$ & $18.586$ & $18.580$ & $18.563$ & $18.516$ & $18.549\pm0.004$ & $0.080\pm0.003$ \\
 & (BVI excl.) &  &  &  &  &  &  &  &  &  & $18.549\pm0.002$ & $0.072\pm0.010$ \\
 &  & $+18$ & $18.807$ & $18.734$ & $18.662$ & $18.535$ & $18.507$ & $18.500$ & $18.506$ & $18.459$ & $18.475\pm0.006$ & $0.083\pm0.005$ \\
 & (BVI excl.) &  &  &  &  &  &  &  &  &  & $18.489\pm0.004$ & $0.046\pm0.020$ \\
 & III & 0 & \nodata & $18.810$ & $18.733$ & $18.615$ & $18.586$ & $18.580$ & $18.563$ & $18.516$ & $18.543\pm0.004$ & $0.089\pm0.005$ \\
 &  & $+18$ & \nodata & $18.748$ & $18.671$ & $18.535$ & $18.507$ & $18.500$ & $18.506$ & $18.459$ & $18.469\pm0.006$ & $0.093\pm0.008$ \\
\hline
\hline
fixed LMC & II & 0 & $18.880$ & $18.781$ & $18.721$ & $18.624$ & $18.594$ & $18.585$ & $18.577$ & $18.527$ & $18.558\pm0.004$ & $0.077\pm0.003$ \\
 & (BVI excl.) &  &  &  &  &  &  &  &  &  & $18.560\pm0.002$ & $0.069\pm0.012$ \\
 &  & $+18$ & $18.794$ & $18.694$ & $18.634$ & $18.538$ & $18.507$ & $18.498$ & $18.513$ & $18.464$ & $18.478\pm0.005$ & $0.074\pm0.004$ \\
 & (BVI excl.) &  &  &  &  &  &  &  &  &  & $18.493\pm0.005$ & $0.042\pm0.028$ \\
 & III & 0 & \nodata & $18.778$ & $18.734$ & $18.624$ & $18.594$ & $18.585$ & $18.577$ & $18.527$ & $18.560\pm0.006$ & $0.075\pm0.007$ \\
 &  & $+18$ & \nodata & $18.691$ & $18.648$ & $18.538$ & $18.507$ & $18.498$ & $18.513$ & $18.464$ & $18.482\pm0.007$ & $0.072\pm0.008$ \\
\hline
unfixed & II & 0 & $18.871$ & $18.778$ & $18.712$ & $18.625$ & $18.589$ & $18.579$ & $18.571$ & $18.518$ & $18.554\pm0.003$ & $0.076\pm0.003$ \\
 & (BVI excl.) &  &  &  &  &  &  &  &  &  & $18.550\pm0.003$ & $0.080\pm0.015$ \\
 &  & $+18$ & $18.785$ & $18.691$ & $18.626$ & $18.539$ & $18.503$ & $18.492$ & $18.507$ & $18.454$ & $18.474\pm0.004$ & $0.074\pm0.004$ \\
 & (BVI excl.) &  &  &  &  &  &  &  &  &  & $18.483\pm0.006$ & $0.053\pm0.031$ \\
 & III & 0 & \nodata & $18.779$ & $18.718$ & $18.625$ & $18.589$ & $18.579$ & $18.571$ & $18.518$ & $18.554\pm0.004$ & $0.076\pm0.005$ \\
 &  & $+18$ & \nodata & $18.693$ & $18.631$ & $18.539$ & $18.503$ & $18.492$ & $18.507$ & $18.454$ & $18.476\pm0.005$ & $0.072\pm0.006$ \\
\hline
fixed MW & II & 0 & $18.874$ & $18.779$ & $18.715$ & $18.624$ & $18.595$ & $18.587$ & $18.578$ & $18.530$ & $18.559\pm0.003$ & $0.075\pm0.002$ \\
 & (BVI excl.) &  &  &  &  &  &  &  &  &  & $18.563\pm0.002$ & $0.066\pm0.011$ \\
 &  & $+18$ & $18.788$ & $18.692$ & $18.629$ & $18.537$ & $18.509$ & $18.500$ & $18.514$ & $18.466$ & $18.480\pm0.004$ & $0.073\pm0.004$ \\
 & (BVI excl.) &  &  &  &  &  &  &  &  &  & $18.495\pm0.005$ & $0.039\pm0.027$ \\
 & III & 0 & \nodata & $18.779$ & $18.717$ & $18.624$ & $18.595$ & $18.587$ & $18.578$ & $18.530$ & $18.561\pm0.003$ & $0.073\pm0.004$ \\
 &  & $+18$ & \nodata & $18.692$ & $18.630$ & $18.537$ & $18.509$ & $18.500$ & $18.514$ & $18.466$ & $18.482\pm0.005$ & $0.069\pm0.006$ \\
\enddata
\tablecomments{Values above double line use a period cutoff of $P>5$ days; values below double line use a period cutoff of $P>10$ days. Blanks indicate that the value is identical to the one above, while an ellipsis (...) indicates no data are available. All errors reported are the standard error of the mean.}
\end{deluxetable*}

\begin{deluxetable*}{ccccccccccrr}

\tablecaption{Distance Moduli and Extinctions from Reddening Law Fits to Samples in the SMC\label{tab:SMC_final}}

\tablehead{\colhead{Slopes (exclusions)} & \colhead{Offset} & \colhead{$\mu_{B}$} & \colhead{$\mu_{V}$} & \colhead{$\mu_{I}$} & \colhead{$\mu_{J}$} & \colhead{$\mu_{H}$} & \colhead{$\mu_{K}$} & \colhead{$\mu_{[3.6]}$} & \colhead{$\mu_{[4.5]}$} & \colhead{$\mu_{0}$} & \colhead{$E(B-V)$} \\ 
\colhead{} & \colhead{($\mu$as)} & \colhead{(mag)} & \colhead{(mag)} & \colhead{(mag)} & \colhead{(mag)} & \colhead{(mag)} & \colhead{(mag)} & \colhead{(mag)} & \colhead{(mag)} & \colhead{(mag)} & \colhead{(mag)}} 

\startdata
fixed LMC OGLE II & 0 & $19.072$ & $19.084$ & $19.067$ & $19.071$ & $19.052$ & $19.056$ & $19.031$ & $18.960$ & $19.049\pm0.004$ & $0.008\pm0.003$ \\
(BVI excl.) &  &  &  &  &  &  &  &  &  & $19.027\pm0.003$ & $0.050\pm0.018$ \\
 & $+18$ & $18.998$ & $19.011$ & $18.994$ & $18.997$ & $18.979$ & $18.983$ & $18.981$ & $18.910$ & $18.983\pm0.003$ & $0.006\pm0.002$ \\
(BVI excl.) &  &  &  &  &  &  &  &  &  & $18.974\pm0.002$ & $0.022\pm0.013$ \\
\hline
fixed LMC OGLE III & 0 & \nodata & $19.085$ & $19.065$ & $19.071$ & $19.052$ & $19.056$ & $19.031$ & $18.960$ & $19.044\pm0.004$ & $0.013\pm0.005$ \\
 & $+18$ & \nodata & $19.012$ & $18.991$ & $18.997$ & $18.979$ & $18.983$ & $18.981$ & $18.910$ & $18.979\pm0.002$ & $0.010\pm0.003$ \\
\hline
unfixed & 0 & $18.804$ & $18.840$ & $18.929$ & $19.009$ & $19.096$ & $19.038$ & $18.883$ & $18.864$ & $19.029\pm0.025$ & $-0.054\pm0.021$ \\
(BVI excl.) &  &  &  &  &  &  &  &  &  & $18.925\pm0.033$ & $0.161\pm0.186$ \\
 & $+18$ & $18.799$ & $18.835$ & $18.924$ & $19.004$ & $19.091$ & $19.033$ & $18.891$ & $18.872$ & $19.028\pm0.023$ & $-0.055\pm0.020$ \\
(BVI excl.) &  &  &  &  &  &  &  &  &  & $18.931\pm0.031$ & $0.146\pm0.176$ \\
\hline
fixed MW & 0 & $19.075$ & $19.082$ & $19.066$ & $19.074$ & $19.058$ & $19.062$ & $19.044$ & $18.973$ & $19.056\pm0.003$ & $0.006\pm0.002$ \\
(BVI excl.) &  &  &  &  &  &  &  &  &  & $19.041\pm0.002$ & $0.037\pm0.014$ \\
 & $+18$ & $18.996$ & $19.005$ & $18.990$ & $18.996$ & $18.980$ & $18.984$ & $18.989$ & $18.918$ & $18.985\pm0.002$ & $0.004\pm0.002$ \\
(BVI excl.) &  &  &  &  &  &  &  &  &  & $18.982\pm0.003$ & $0.010\pm0.014$ \\
\hline
\hline
fixed LMC OGLE II & 0 & $19.069$ & $19.074$ & $19.061$ & $19.049$ & $19.040$ & $19.047$ & $19.055$ & $18.978$ & $19.045\pm0.002$ & $0.007\pm0.002$ \\
(BVI excl.) &  &  &  &  &  &  &  &  &  & $19.052\pm0.003$ & $-0.009\pm0.014$ \\
 & $+18$ & $18.982$ & $18.987$ & $18.974$ & $18.963$ & $18.953$ & $18.960$ & $18.992$ & $18.914$ & $18.966\pm0.005$ & $0.005\pm0.004$ \\
(BVI excl.) &  &  &  &  &  &  &  &  &  & $18.985\pm0.006$ & $-0.036\pm0.032$ \\
\hline
fixed LMC OGLE III & 0 & \nodata & $19.072$ & $19.066$ & $19.049$ & $19.040$ & $19.047$ & $19.055$ & $18.978$ & $19.044\pm0.002$ & $0.009\pm0.003$ \\
 & $+18$ & \nodata & $18.985$ & $18.980$ & $18.963$ & $18.953$ & $18.960$ & $18.992$ & $18.914$ & $18.965\pm0.005$ & $0.006\pm0.006$ \\
\hline
unfixed & 0 & $18.350$ & $18.635$ & $18.759$ & $18.729$ & $18.849$ & $18.814$ & $18.937$ & $18.842$ & $18.910\pm0.025$ & $-0.116\pm0.021$ \\
(BVI excl.) &  &  &  &  &  &  &  &  &  & $18.963\pm0.018$ & $-0.258\pm0.102$ \\
 & $+18$ & $18.268$ & $18.553$ & $18.677$ & $18.647$ & $18.767$ & $18.732$ & $18.870$ & $18.776$ & $18.833\pm0.026$ & $-0.118\pm0.021$ \\
(BVI excl.) &  &  &  &  &  &  &  &  &  & $18.894\pm0.020$ & $-0.276\pm0.110$ \\
\hline
fixed MW & 0 & $19.067$ & $19.072$ & $19.055$ & $19.050$ & $19.036$ & $19.042$ & $19.051$ & $18.972$ & $19.042\pm0.002$ & $0.007\pm0.002$ \\
(BVI excl.) &  &  &  &  &  &  &  &  &  & $19.045\pm0.003$ & $-0.001\pm0.017$ \\
 & $+18$ & $18.981$ & $18.986$ & $18.968$ & $18.963$ & $18.950$ & $18.955$ & $18.987$ & $18.908$ & $18.962\pm0.004$ & $0.005\pm0.004$ \\
(BVI excl.) &  &  &  &  &  &  &  &  &  & $18.978\pm0.006$ & $-0.028\pm0.035$ \\
\enddata
\tablecomments{See note to Table \ref{tab:LMC_final}}
\end{deluxetable*}

The fit in best agreement with DEBs uses all available wavelength-dependent distance moduli ($BVI_{c}$, $JHK_{s}$, and $[3.6]$) resulting from PL relations fixed to the LMC slopes, including a +18 $\mu$as offset to the Milky Way parallaxes, as discussed in Section \ref{subsec:plx_off} below. This fit is shown in Figure \ref{fig:final_DMs}. We find distances and corresponding standard errors of the means of $\mu_{0} = 18.472 \pm 0.004$ mag to the LMC and $\mu_{0} = 18.983 \pm 0.002$ mag to the SMC. We find a total line-of-sight reddening of $E(B-V) = 0.081 \pm 0.007$ mag to the LMC and $E(B-V) = 0.006 \pm 0.005$ mag to the SMC. We note that other samples of stars may have different line-of-sight reddenings.

We report the wavelength-dependent apparent distance moduli along with the true distance moduli and reddenings for various reddening curve fits to the LMC and SMC in Tables \ref{tab:LMC_final} and \ref{tab:SMC_final} respectively. There are a few notable features. Results for the three slope-fitting methods (fixed to the LMC, fixed to the Milky Way, and unfixed) are similar in all cases, as are the respective fits using only the $JHK_{s}$ and $[3.6]$ micron bands. There is also no significant difference between using a lower period cutoff of $\log P > 0.7$ versus a cutoff of $\log P > 1.0$, which is the more commonly adopted cutoff in distant extragalactic distance measurements. \cite{2009ApJ...693..691N} and \cite{2004A&A...424...43S} found evidence that the Cepheid PL relation has some non-linearities below a period of 10 days; however, the consistency between results with different period cutoffs indicates that our results are not significantly affected by this. The largest differences are driven by whether or not the +18 $\mu$as offset is applied, as expected.

\medskip\medskip
\subsection{Wesenheit Distance Moduli\label{sec:Wes_DMs}}

We also determine the distance moduli to both galaxies using multiple formulations of the reddening-free Wesenheit function, introduced in \cite{1982ApJ...253..575M}. This function eliminates reddening based on an assumed reddening law and coefficient. We use three formulations: an optical function (Equation \ref{eqn:1}), a near-infrared function (Equation \ref{eqn:2}), and a composite three-band function (Equation \ref{eqn:3}) as constructed in \cite{2011ApJ...730..119R, 2016ApJ...826...56R, 2021ApJ...908L...6R}. We use the reddening law and $R_{V}$ value described in Section \ref{subsec:Gaia_MW_dists} to calculate the coefficients for the color term of each function. Note that $W_{H,\,VI}$ is only calculable for the SMC and Milky Way samples, as the optical and near-infrared samples for the LMC do not significantly overlap. 

\begin{equation}\label{eqn:1}
    W_{VI} = V - 2.61 (V - I)
\end{equation}

\begin{equation}\label{eqn:2}
    W_{JK} = J - 1.71 (J - K)
\end{equation}

\begin{equation}\label{eqn:3}
    W_{H,\,VI} = H - 0.496 (V - I)
\end{equation}

\begin{figure*}
    \centering
    \includegraphics[width=\textwidth]{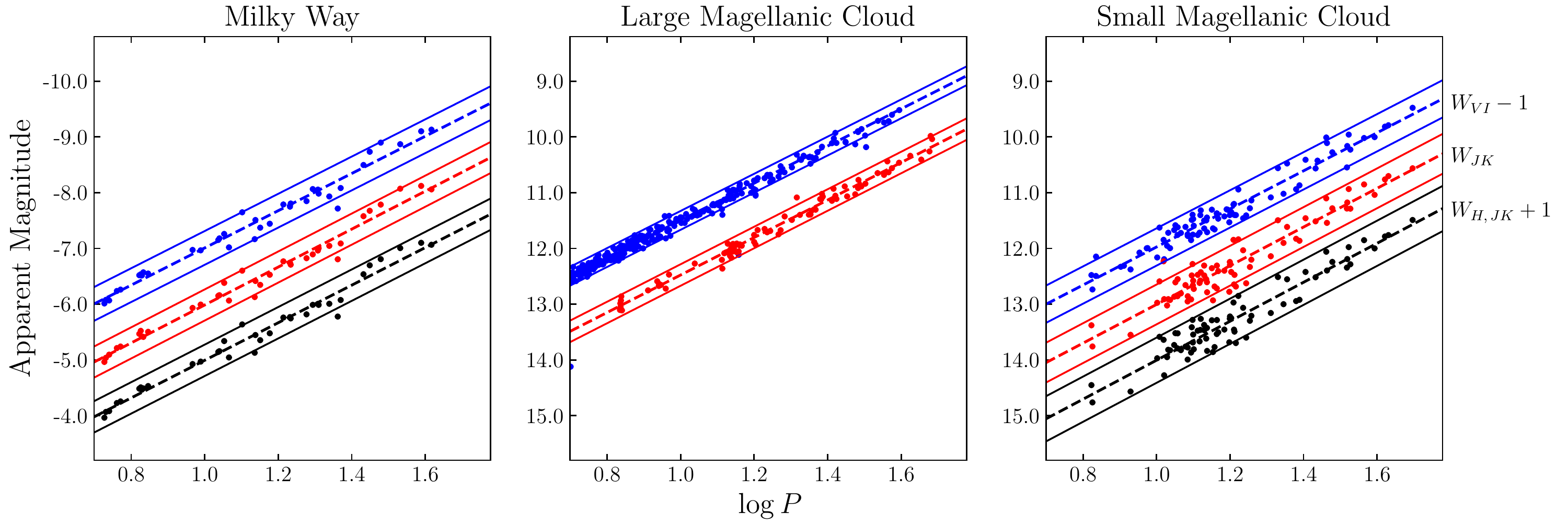}
    \caption{Unfixed-slope PL relations for the three formulations of the Wesenheit function in each galaxy. The LMC does not have a PL relation for $W_{H,\,VI}$ because the optical and near-infrared samples do not significantly overlap. Note that the LMC scatter is very small, compared to the scatter in the Milky Way and SMC. Large scatter is expected in the SMC due to the back-to-front geometry; however, the large scatter in the Milky Way indicates that parallax uncertainties may be a significant source of scatter.}
    \label{fig:W_PLR}
\end{figure*}

We show the Wesenheit PL relations for the three galaxies in Figure \ref{fig:W_PLR}, and the corresponding equations, scatter, and distance moduli are given in Table \ref{tab:W_PLRs}, including a +18 $\mu$as offset for the Milky Way Cepheids. The results are largely consistent with the analysis based on the full reddening curve fit described in Section \ref{subsec:EDR3dist}, indicating that our issue is unlikely to be the individual reddenings to Galactic Cepheids.

The scatter in the Wesenheit PL relations for the Milky Way is notably larger than expected. The SMC PL relations are known to have a large scatter due to the SMC's back-to-front geometry \citep{2016ApJ...816...49S}, but the Milky Way Cepheids should not have this issue since their distances have been individually determined. In Section \ref{subsec:sys_met}, we constrain the size of the metallicity effect using DEBs, finding that metallicity effects are small and therefore not likely the cause of the large scatter. Excess scatter beyond the astrophysically-driven scatter resulting from the intrinsic width of the instability strip and the metallicity differences are thus likely driven by parallax errors in the Milky Way data. Reddening errors also cannot be responsible for the excess Milky Way PL scatter since the Wesenheit functions remove reddening implicitly.

\begin{deluxetable}{c|lcrcc}
\tablecaption{Wesenheit Period-Luminosity Relations}\label{tab:W_PLRs}
\tablehead{\colhead{} & \colhead{Filter} & \colhead{a} & \colhead{b} & \colhead{$\sigma$} & \colhead{$(m-M)_{0}$}} 
\startdata
\multirow{3}{*}{MW} & $W_{VI}$ & $-3.339\pm0.099$ & $-6.677\pm0.025$ & 0.151 & \nodata \\
 & $W_{JK}$ & $-3.401\pm0.091$ & $-6.666\pm0.023$ & 0.140 & \nodata \\
 & $W_{H,\,VI}$ & $-3.370\pm0.092$ & $-6.667\pm0.023$ & 0.141 & \nodata \\
\hline
\multirow{2}{*}{LMC} & $W_{VI}$ & $-3.333\pm0.069$ & $11.830\pm0.005$ & 0.084 & 18.507 \\
 & $W_{JK}$ & $-3.363\pm0.048$ & $11.804\pm0.011$ & 0.096 & 18.470 \\
\hline
\multirow{3}{*}{SMC} & $W_{VI}$ & $-3.415\pm0.094$ & $12.291\pm0.018$ & 0.167 & 18.968 \\
 & $W_{JK}$ & $-3.472\pm0.140$ & $12.310\pm0.026$ & 0.224 & 18.976 \\
 & $W_{H,\,VI}$ & $-3.500\pm0.134$ & $12.306\pm0.025$ & 0.213 & 18.973 \\
\enddata

\end{deluxetable}

We adopt the LMC scatter as our best approximation of the intrinsic scatter in the Wesenheit functions, though we note that this assumes no additional scatter due to back-to-front geometry/tilt effects in the LMC. We use this value to estimate the scatter due to parallax errors in the Milky Way Cepheid sample. The average of the Wesenheit scatter for the different waveband combinations in the LMC is $\sigma_{\mathrm{LMC}} = \pm 0.090$ mag and in the Milky Way the scatter is $\sigma_{\mathrm{MW}} = \pm 0.144$ mag. Assuming that scatter adds in quadrature, we calculate the excess scatter due to parallax errors in the Milky Way sample to be $\sqrt{0.144^{2} - 0.090^{2}} = \pm 0.112$ mag. This does not account for differences in internal scatter from metallicity. However, \cite{2008A&A...488..731R} find the range of measured metallicites for 21 Cepheids in the LMC is 0.51 mag. The range of metallicities for 37 Cepheids in the Milky Way sample is 0.67 mag, so any metallicity induced scatter should be similarly represented in both samples.
\medskip\medskip
\section{EDR3 Bright Cepheid Data Quality}\label{sec:Qual}

The precision and accuracy of \textit{Gaia} parallaxes have increased considerably since the first data release (DR1). Though \textit{Gaia} marked a revolutionary improvement from the \textit{Hipparcos} era, the DR1 parallaxes were given only for the couple million stars in common with the Tycho-2 catalog, using a Bayesian prior on the distances. These parallaxes carried statistical errors of 0.3 mas on average, in addition to a 0.3 mas systematic-error component \citep{2016A&A...595A...2G}, and independent analyses found zero-point offsets upward of a few hundred $\mu$as \cite[e.g.][]{2016ApJ...831L...6S, 2016ApJ...832L..18J}. Two years later, the second data release expanded parallax measurements to cover over 1.3 \textit{billion} sources, and for sources brighter than $G=15$ mag, average statistical and systematic errors were decreased to below 0.07 and 0.10 mas, respectively. The average parallax offset decreased ten-fold to $-29$ $\mu$as as determined from quasars in \cite{2018A&A...616A...2L}. However, trends in offset, based on magnitude, color, and position on the sky, were identified, so the \textit{Gaia} Collaboration recommended treating the zero point as an adjustable parameter in fitting. Further, the treatment of bright stars was problematic, creating the possibility of a different average parallax offset for $G < 13$ mag.

Now, utilizing nearly three years of observations, EDR3 has reduced the average statistical errors to below 0.02 mas for $G \leq 15$ sources, making systematic errors a significant portion of the error budget. \cite{2020arXiv201201742L} (hereafter L21b) characterized the parallax zero point globally, fitting a dependence on magnitude, color/chromaticity, and galactic latitude. They find the average parallax offset has been reduced to $-17$ $\mu$as in EDR3, compared to -29 $\mu$as in DR2. The prescription was calculated from 1.1 million quasars and sources in the LMC for $G>13$, and extended to brighter sources ($6<G<13$) using physical pairs (resolved binaries). The solution is well-characterized for intermediate colors, described by the effective wavenumber $1.24 < \nu_{\mathrm{eff}} < 1.72$. However, corrections for very bright sources are significantly more uncertain, due to the small number of available calibrators.

Bright-star astrometry presents a challenge for \textit{Gaia}. EDR3 reports $G$ magnitudes spanning roughly 5 to 20 mag, which is a factor of \textit{a million} in brightness. This dynamic range cannot be achieved using simple integration, as detector saturation begins to occur at $G \simeq 12$ mag, complicating the source centroiding. To mitigate this, the telescope utilizes a complex system of windows and gates to minimize saturation. Light from a source is first passed through the ``Sky Mapper" which assigns it a particular window based on an initial flux estimate, and if necessary time-delayed integration (TDI) gates are activated to decrease the exposure time and further prevent saturation effects. The activation of gating around $G=12.5$~mag creates a discontinuity in the parallax offsets, so stars at this boundary may have additionally uncertain parallaxes. Further, for very bright stars, where gating can no longer prevent saturation, the point spread function (PSF) models cannot fully fit stars for which the entire core region of the PSF is saturated. Thus, there is a threshold in the bright stars after which noise is introduced by the saturation, and several of our Cepheids may well be above this threshold.

In Section \ref{subsec:plx_off}, we describe our efforts to determine a parallax offset for our bright Cepheid sample directly from the EDR3 data, and in Section \ref{subsec:DEB_comp}, we derive an offset by comparing to previous distances to the LMC and SMC derived from detached eclipsing binaries. In Section \ref{subsec:HST_comp}, we compare the EDR3 parallaxes for individual Cepheids to previous distances from HST.

\subsection{Parallax Offset}\label{subsec:plx_off}

\begin{figure}
    \centering
    \includegraphics[width=\columnwidth]{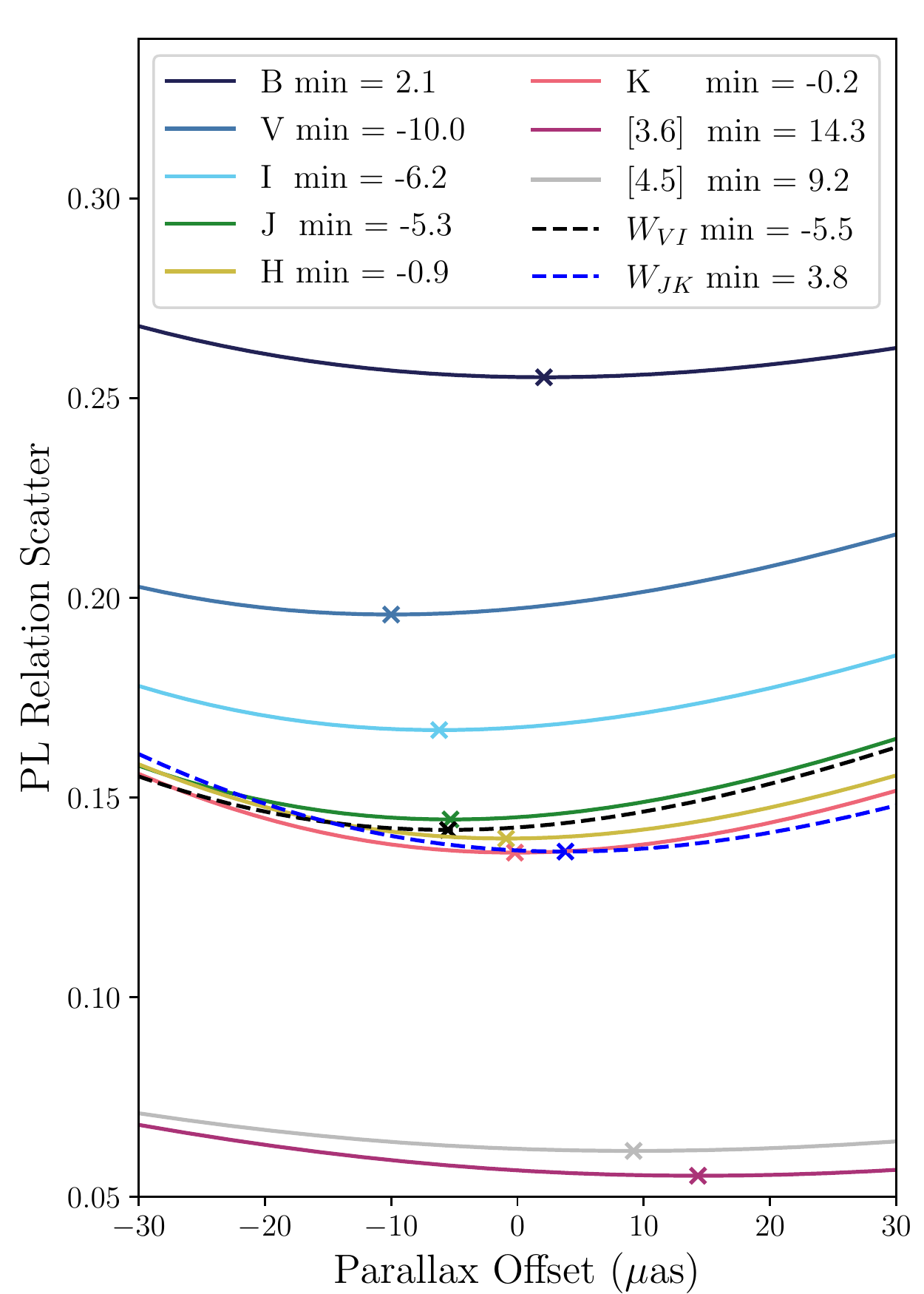}
    \caption{Scatter in the PL relation for each band, including two formulations of the Wesenheit function, vs. the parallax offset. The ``x" points mark the location of minimum scatter. There is not a universal minimum across bands, and the Wesenheit functions' minima also disagree, indicating that the disagreement is not a reddening-induced effect. \label{fig:MW_offset}}
\end{figure}

Many recent publications have speculated on the existence of a parallax offset that is in addition to the \textit{Gaia} Collaboration's prescribed zero-point, particularly for bright stars. This was first suggested in \cite{2021ApJ...908L...6R} (hereafter R21), who observed Cepheids in a similar magnitude and color range to our sample. They derived an offset of  $-14 \pm 6$ $\mu\textrm{as}$ (meaning that the parallaxes were \textit{overcorrected} by the L21b offset) using a chi-squared minimization of their ``photometric parallaxes." Their preferred two-parameter fit used fixed slope and metallicity scalings, independently fitting only for the intercept of the PL relation at $P=10$ days and for the applicable EDR3 zero-point offset. They find that a $-14$ $\mu$as offset best minimizes their PL relation scatter in $W$, adopting a $\pm$10 $\mu$as prior on the uncertainty on the L21b correction, or a $-17 \pm 6$ $\mu$as offset if they do not adopt an uncertainty prior.

Shortly thereafter, \cite{2021arXiv210107252Z} derived a similar offset of $-15 \pm 3$ $\mu\textrm{as}$ from first-ascent red-giant branch stars with $G<10.8$ mag, using asteroseismetric distances from APOKASC-2. A calibration error in asteroseimetric distances results in a difference from the geometric distances that scales with the parallax, so they fit for both a parallax dependent term to account for these errors and a constant offset for the EDR3 parallaxes. For stars with $G>10.8$ mag, they do not find evidence for a significant offset, indicating it may be purely a bright star issue. \cite{2021arXiv210109691H} find a similar result for the LAMOST primary red clump (PRC) stellar sample, with an offset of $-9.8 \pm 1.0$ $\mu\textrm{as}$ for $<10.8$ mag, although they also find an offset of $-9.0 \pm 0.4$ $\mu\textrm{as}$ for $G>14$. The sample as a whole ($9<G<15$) is found to confirm the L21b offset, with residuals on the order of a few $\mu$as. \cite{2021arXiv210110206M} find a smaller offset in the opposite direction of $+6.9 \pm 2.2$ $\mu\textrm{as}$ using globular clusters for stars in the range $9.3<G<11$ mag.

Although the publications cited above found significant offsets from the L21b-corrected EDR3 parallaxes, this is not the case for all studies, even at bright $G$ magnitudes. \cite{2021ApJ...907L..33S} investigated eclipsing binaries from $5<G<12$ mag, precisely in the Cepheid range, and found a mean additional offset of $+15 \pm 18$ $\mu\textrm{as}$, which is consistent with zero at the less than one-sigma level. However, we should note that based on their sample size, $\pm$15 $\mu\textrm{as}$ was the maximum possible precision, which would be insufficient to determine an offset on the small scales of the aforementioned studies. \cite{2021arxiv210811391R} found a $+15\pm15$ $\mu$as offset by comparing 7 of the Cepheids with HST scanning parallaxes from \cite{2018ApJ...855..136R} to the EDR3 parallaxes, though they emphasize that this is a small sample where two of the stars have large scatter with respect to the rest of the sample. Their overall result indicates the L21b offsets sufficiently correct the parallaxes within the quoted  uncertainties. \cite{2021arXiv210310894B} derive a metallicity dependence using the Magellanic Clouds Cepheids in combination with EDR3 data for Galactic Cepheids, without the need for any additional offset. \cite{2021arXiv210302077G} also calibrate the surface-brightness/color relations of bright detached eclipsing binaries ($5.5<G<12$~mag) in the color range $1.45<\nu_{\mathrm{eff}}<1.75$, also without invoking an additional parallax offset.

We investigated the existence of a parallax offset in our sample, as our initially derived distances were much larger than previously determined distances to the Magellanic Clouds, as shown in Table \ref{tab:DM_literature}. Thus, we attempted to fit the offset based on our own EDR3 data, and as in the R21 study, we looked for minima in the scatter of the PL relations. However, because we do a multi-wavelength fit, we did not simply constrain ourselves to using a single bandpass, but we fit for the minimum in each wavelength. Distance errors are achromatic, and for a given star they propagate identically in all wavelengths, as the correcting distance term on magnitude is the same. Longer wavelengths have decreasing intrinsic scatter due to the decreased sensitivity of the surface brightness to temperature, so the fractional contribution from the distance term increases. However, removing a constant distance error should still monotonically reduce scatter in all bands. We would expect to see that the scatter would reach a minimum at a similar offset value, allowing some flexibility for reddening errors or other issues.

\begin{deluxetable}{c|ccc}
\label{tab:DM_literature}
\tablecaption{Compiled Distance Moduli and Extinctions}

\tablehead{\colhead{} & \colhead{$\mu_{0}$} & \colhead{$E(B-V)$} & \colhead{Reference}} 

\startdata
\multirow{6}{*}{LMC} & $18.477 \pm 0.030$ & \nodata & \cite{2019Natur.567..200P} \\
 & $18.48 \pm 0.03$ & $0.12 \pm 0.01$ & \cite{2012ApJ...759..146M} \\
 & $18.50 \pm 0.03$ & \nodata & \cite{2007AJ....133.1810B} \\
 & \nodata & $0.091\pm0.050$ & \cite{2019AA...628A..51J} \\
 & $18.538\pm0.063$ & $0.083\pm0.028$ & this work (no offset) \\
 & $18.472\pm0.091$ & $0.081\pm0.028$ & this work (+18 $\mu$as offset) \\
 \hline
\multirow{5}{*}{SMC} & $18.977 \pm 0.044$ & \nodata & \cite{2020ApJ...904...13G} \\
 & $18.96 \pm 0.04$ & $0.071 \pm 0.004$ & \cite{2016ApJ...816...49S} \\
 & \nodata & $0.038\pm0.053$ & \cite{2019AA...628A..51J} \\
 & $19.050\pm0.076$ & $0.008\pm0.034$ & this work (no offset) \\
 & $18.983\pm0.101$ & $0.006\pm0.034$ & this work (+18 $\mu$as offset) \\
\enddata
\end{deluxetable}

\begin{deluxetable}{lcccc}

\tablecaption{Parallax Offset Distance Moduli and Extinctions}\label{tab:offsets}

\tablehead{\colhead{Offset} & \colhead{$(m-M)^{\mathrm{LMC}}_{0}$} & \colhead{$E(B-V)^{\mathrm{LMC}}$} & \colhead{$(m-M)^{\mathrm{SMC}}_{0}$} & \colhead{$E(B-V)^{\mathrm{SMC}}$} \\ 
\colhead{($\mu$as)} & \colhead{(mag)} & \colhead{(mag)} & \colhead{(mag)} & \colhead{(mag)} } 

\startdata
0.0 & $18.540\pm0.004$ & $0.082\pm0.007$ & $19.051\pm0.004$ & $0.008\pm0.005$ \\
3.0 & $18.529\pm0.004$ & $0.082\pm0.007$ & $19.039\pm0.003$ & $0.007\pm0.005$ \\
5.0 & $18.521\pm0.004$ & $0.082\pm0.007$ & $19.032\pm0.003$ & $0.007\pm0.005$ \\
7.0 & $18.513\pm0.004$ & $0.082\pm0.007$ & $19.024\pm0.003$ & $0.007\pm0.005$ \\
10.0 & $18.502\pm0.004$ & $0.081\pm0.007$ & $19.013\pm0.003$ & $0.006\pm0.005$ \\
13.0 & $18.491\pm0.004$ & $0.081\pm0.007$ & $19.001\pm0.003$ & $0.006\pm0.005$ \\
15.0 & $18.483\pm0.004$ & $0.080\pm0.007$ & $18.994\pm0.002$ & $0.006\pm0.005$ \\
16.0 & $18.480\pm0.004$ & $0.080\pm0.007$ & $18.990\pm0.002$ & $0.005\pm0.005$ \\
17.0 & $18.476\pm0.004$ & $0.080\pm0.007$ & $18.987\pm0.002$ & $0.005\pm0.005$ \\
18.0 & $18.472\pm0.004$ & $0.080\pm0.007$ & $18.983\pm0.002$ & $0.005\pm0.005$ \\
20.0 & $18.465\pm0.004$ & $0.080\pm0.007$ & $18.975\pm0.002$ & $0.005\pm0.005$ \\
23.0 & $18.454\pm0.004$ & $0.079\pm0.008$ & $18.964\pm0.002$ & $0.005\pm0.005$ \\
25.0 & $18.446\pm0.004$ & $0.079\pm0.008$ & $18.957\pm0.003$ & $0.004\pm0.005$ \\
27.0 & $18.439\pm0.005$ & $0.079\pm0.008$ & $18.950\pm0.003$ & $0.004\pm0.005$ \\
30.0 & $18.428\pm0.005$ & $0.079\pm0.008$ & $18.939\pm0.003$ & $0.004\pm0.005$ \\
\enddata
\end{deluxetable}

Unfortunately, this is not found in the data. Figure \ref{fig:MW_offset} shows minima ranging from $-15$ to $+15$ $\mu$as for the different bandpasses. The minima decrease roughly as a function of wavelength. We have calculated the minima of the two Wesenheit magnitudes given in Equations \ref{eqn:1} and \ref{eqn:2} to make sure that reddening issues are not the source of the the discrepancy. This does not solve the issue, as the two reddening-free measures disagree with each other, with the minimum in $W_{JK}$ being +3.8 $\mu$as and the minimum in $W_{VI}$ being -5.5 $\mu$as. Additionally, these values are not close to the values of +9.2 and +14.3 $\mu$as in the mid-IR, where reddening is negligible. Therefore, we conclude that the parallax offset cannot be self-consistently calibrated with the data currently available to us.

As a whole, it is clear that the existence of an additional parallax offset is still being debated; no agreed-upon value exists for the bright stars, and even the sign of the offset varies from sample to sample. Further, we note that determining an offset for Cepheid samples is particularly difficult due to uncertainty about the effect of metallicity on the intercept of the PL relation. The value of the additional offset is directly covariant with the value of any applied metallicity correction if comparing to external distances. This could potentially be mitigated by using high-quality multi-wavelength data; however, this would be dependent upon knowing the values of the metallicity corrections to very high accuracy and precision. As we discuss in Section \ref{subsec:sys_met}, this is not currently the case. This covariance presents a significant barrier to determining the high precision distances necessary for anchoring 1\% measurements of $H_{0}$.

\subsection{Comparison with the DEB Distances}\label{subsec:DEB_comp}

\begin{figure*}
    \centering
    \includegraphics[width=450pt]{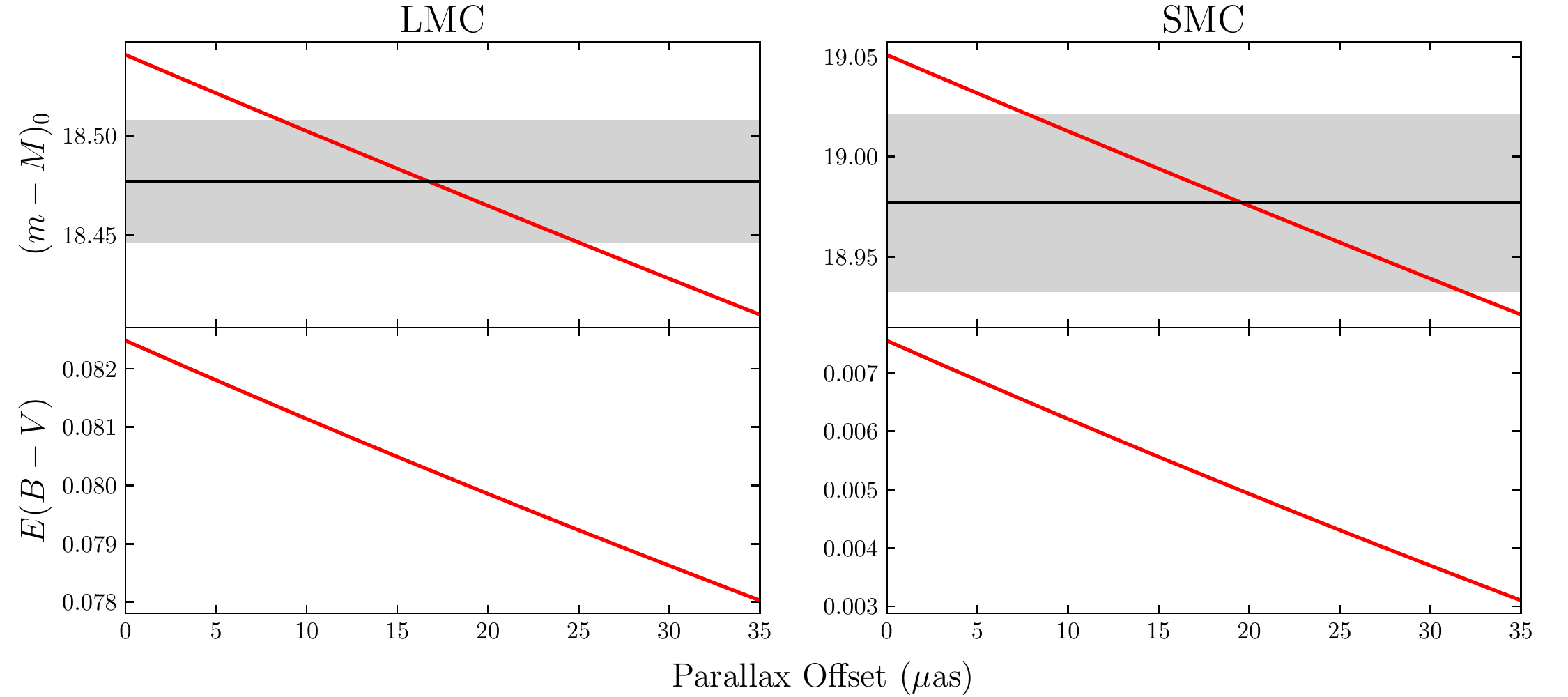}
    \caption{Distance moduli (top panel) and extinctions (bottom panel) shown in red for values of the parallax offset ranging from 0 to +35 $\mu$as for the LMC and SMC. The black line shows the detached eclipsing binary distance modulus with the 1$\sigma$ errors shaded in gray. The offsets in best agreement with the DEB distances are +17 $\mu$as for the LMC and +20 $\mu$as for the SMC.}
    \label{fig:DEB_off}
\end{figure*}

Since we were unable to determine the offset directly from the data, we have instead calculated an offset by incorporating external data sets for which accurate geometric measurements are available. We did this by comparing our derived distance moduli to the LMC and SMC directly with the most accurate measurements as given by DEBs \citep{2019Natur.567..200P, 2020ApJ...904...13G}, with uncertainties of 1\% and 2\%, respectively.

We calculated distance moduli by applying fixed slopes from the LMC to the other data sets at all available wavelengths ($B$ through $[3.6]$). Next we recalculated these fits for different linear offsets in the range 0 to +30 $\mu$as. The trajectories in both distance modulus and derived color excess are shown in Figure \ref{fig:DEB_off} and calculated explicitly for several values in Table \ref{tab:offsets}. We show the error bars on the DEB measurements based on their summed statistical and systematic errors. From these fits, we derive a mean offset of $+17 \pm 8~\mu$as for the LMC, and $+20 \pm 12~\mu$as for the SMC. Taking an error-weighted average results in an offset value of $+18\pm14~\mu$as, summing the errors in quadrature. We determine distance moduli both with and without this offset, given in Tables \ref{tab:LMC_final} and \ref{tab:SMC_final}.

\subsection{HST Parallax Comparison}\label{subsec:HST_comp}

\begin{figure}
    \centering
    \includegraphics[width=\columnwidth]{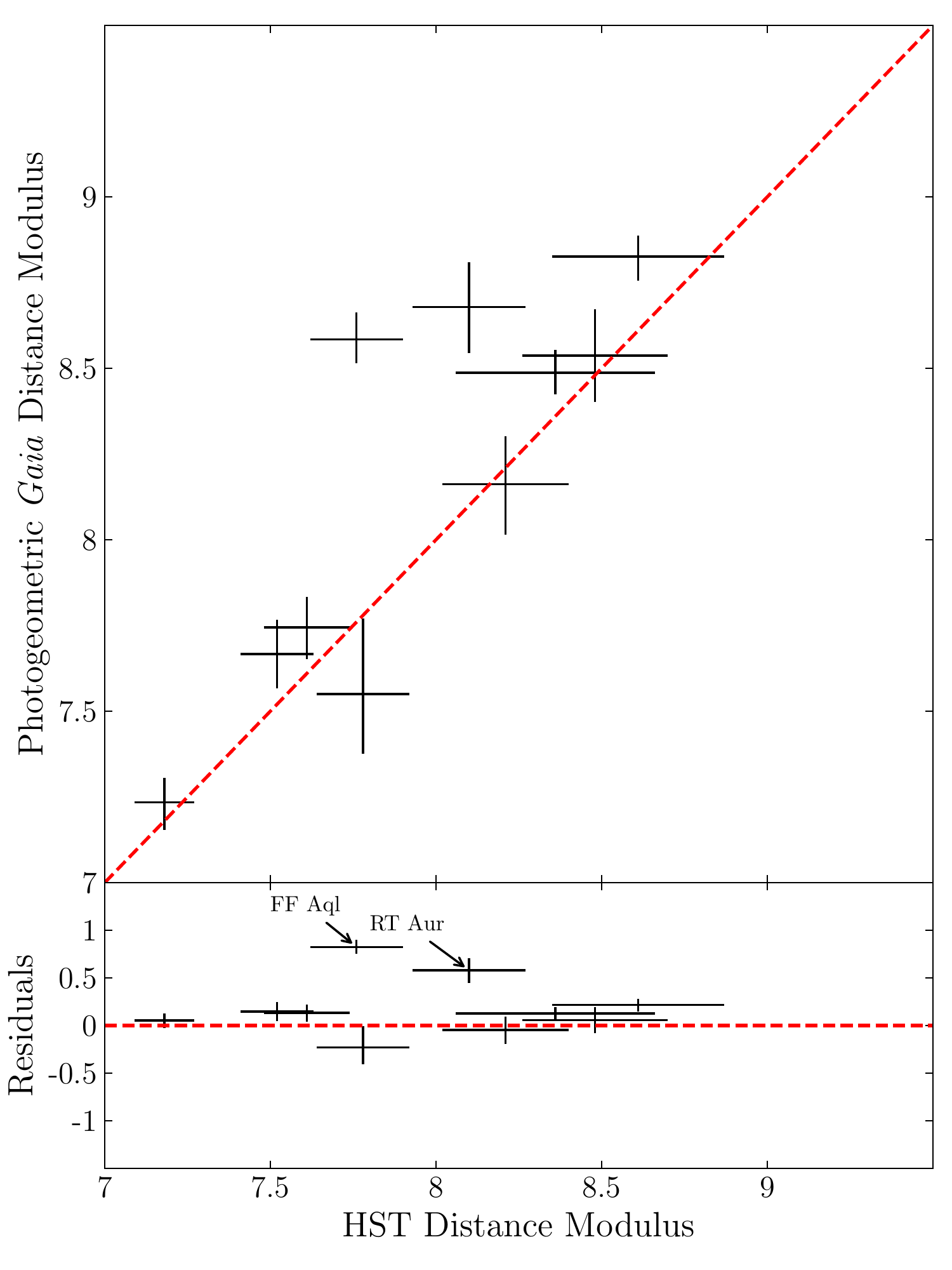}
    \caption{Comparison between distance moduli from HST fine-guidance sensor parallaxes and from \textit{Gaia} EDR3 parallaxes for ten nearby Cepheids. Two Cepheids (FF Aql and RT Aur) have significantly different distances between the two samples.}
    \label{fig:Ben_comp}
\end{figure}

Ten of the nearest Cepheids in our sample have HST trigonometric parallaxes, measured by \cite{2002AJ....124.1695B, 2007AJ....133.1810B}. These stars are very bright on ensemble, with a mean apparent $G$ magnitude of $4.5\pm0.8$ mag, compared to the full quality-selected sample of 37 Cepheids which has a mean apparent $G$ magnitude of $7.7\pm1.4$ mag. The average reported error on the HST parallaxes is 8.0\% while the average reported error on the corresponding \textit{Gaia} EDR3 parallaxes is 5.1\%. We directly compare the distances derived from these two measurements in Figure \ref{fig:Ben_comp}.

Notably, two Cepheids (FF Aql and RT Aur) show significant differences in the distances derived from the two independent parallax determinations. RT Aur has {\tt ruwe} $=6.44$, so this star was flagged as low quality by our quality cuts. However, FF Aql has a {\tt ruwe} value of 1.06. It was not included in the final sample due to its short period, but we note that it would not have otherwise been removed by standard cuts. Given the issues for \textit{Gaia} measurements of bright stars described earlier, it may also be the case that the qualities of some bright stars are also overestimated, and should be treated with care. Thus, in some cases a simple goodness-of-fit cut may not be sufficient to ensure the quality of bright samples. For this reason, we have adopted conservative errors overall for the final distances to the LMC and SMC, derived from these parallaxes, as described in Section \ref{sec:err_bud} below.

In addition, there is an offset between the HST and EDR3 parallaxes, as expected from our need for an average parallax offset to bring the distance moduli into concordance with those based on DEBs. Eight of the ten Cepheids in this sample have larger distance moduli according to \textit{Gaia} than HST. It is not clear if this is purely a result of random parallax errors or if there is a systematic offset of the parallaxes in this bright regime. Based solely on this sample of ten Cepheids, the scatter is consistent with being due to random parallax errors since the scatter is larger than the measured offset, and using only the HST parallax sample as a comparison would result in a parallax offset of +186 $\mu$as, an order of magnitude larger than the offset derived from DEBs. If the two outliers are excluded, this would still be an average offset of +57 $\mu$as. Thus, we remain cautious in interpreting the ``parallax offset" term discussed in Sections \ref{subsec:plx_off} and \ref{subsec:DEB_comp} as a universal quantity, as it may be quite sample-specific.

\section{Error Budget}\label{sec:err_bud}

We divide the overall error budget into two types of errors: reducible and irreducible. Reducible errors include any type of statistical errors which can be mitigated with larger sample size, as well as errors which result from systematic qualities of our particular sample. Irreducible errors are errors which cannot be decreased significantly by averaging over larger samples nor by constructing more robust samples.

In Section \ref{subsec:bootstrap}, we calculate the statistical errors on the distance moduli and reddenings to the LMC and SMC. In Sections \ref{subsec:sys_zpL20}, \ref{subsec:sys_met}, \ref{subsec:sys_red}, and \ref{subsec:sys_geom} we calculate systematic errors on these same quantities due to metallicity effects, the parallax zero-point and offset, reddening corrections, and the geometry of the Magellanic Clouds. Our final error budget is given in Table \ref{tab:budget}, and the irreducible systematic error of the EDR3-calibrated Cepheid distance relation is tabulated in Table \ref{tab:sys_bud}.

\begin{deluxetable*}{c|cc|cc|cc|cc}
\tablewidth{0pt}
\tablecaption{Error Budget for LMC and SMC Distances and Reddenings from Reddening Law Fit\label{tab:budget}}
\tablecolumns{9}
\tablehead{
\multicolumn{1}{c}{ } &
\multicolumn{2}{c}{$\mu_{0, \,\mathrm{LMC}}$} &
\multicolumn{2}{c}{$E(B-V)_{\mathrm{LMC}}$} &
\multicolumn{2}{c}{$\mu_{0, \,\mathrm{SMC}}$} &
\multicolumn{2}{c}{$E(B-V)_{\mathrm{SMC}}$} \\
\colhead{Source of Uncertainty [mag]} & 
\colhead{$\sigma_{stat}$} & 
\colhead{$\sigma_{sys}$} &
\colhead{$\sigma_{stat}$} & 
\colhead{$\sigma_{sys}$} & 
\colhead{$\sigma_{stat}$} & 
\colhead{$\sigma_{sys}$} &
\colhead{$\sigma_{stat}$} & 
\colhead{$\sigma_{sys}$}
}
\startdata
Bootstrapping Analysis (Section \ref{subsec:bootstrap}) & 0.024 & \nodata & 0.012 & \nodata & 0.032 & \nodata & 0.013 & \nodata \\
Metallicity Effect (Section \ref{subsec:sys_met})& \nodata & 0.028 & \nodata & \nodata & \nodata & 0.053 & \nodata & \nodata \\
Parallax Zero Point Correction (L21b) & \nodata & 0.037 & \nodata & 0.001 & \nodata & 0.037 & \nodata & 0.001 \\
Parallax Zero Point Offset (Section \ref{subsec:sys_zpL20})& \nodata & 0.054 & \nodata & 0.002 & \nodata & 0.054 & \nodata & 0.002 \\
Reddenings Adjustment (Section \ref{subsubsec:red_adj}) & \nodata & \nodata & \nodata & 0.014 & \nodata & \nodata & \nodata & 0.014 \\
$R_{V}$ Variation (Section \ref{subsubsec:sys_red_coeff}) & \nodata & 0.002 & \nodata & 0.006 & \nodata & 0.002 & \nodata & 0.001 \\
Magellanic Clouds Geometry (Section \ref{subsec:sys_geom}) & \nodata & 0.005 & \nodata & \nodata & \nodata & 0.025 & \nodata & \nodata \\
\hline
\hline
Cumulative Errors (mag)& 0.024 & 0.071 & 0.012 & 0.014 & 0.032 & 0.087 & 0.013 & 0.014 \\
\enddata
\end{deluxetable*}

\subsection{Statistical Error From Bootstrapping\label{subsec:bootstrap}}

\begin{figure*}
    \centering
    \includegraphics[width=450pt]{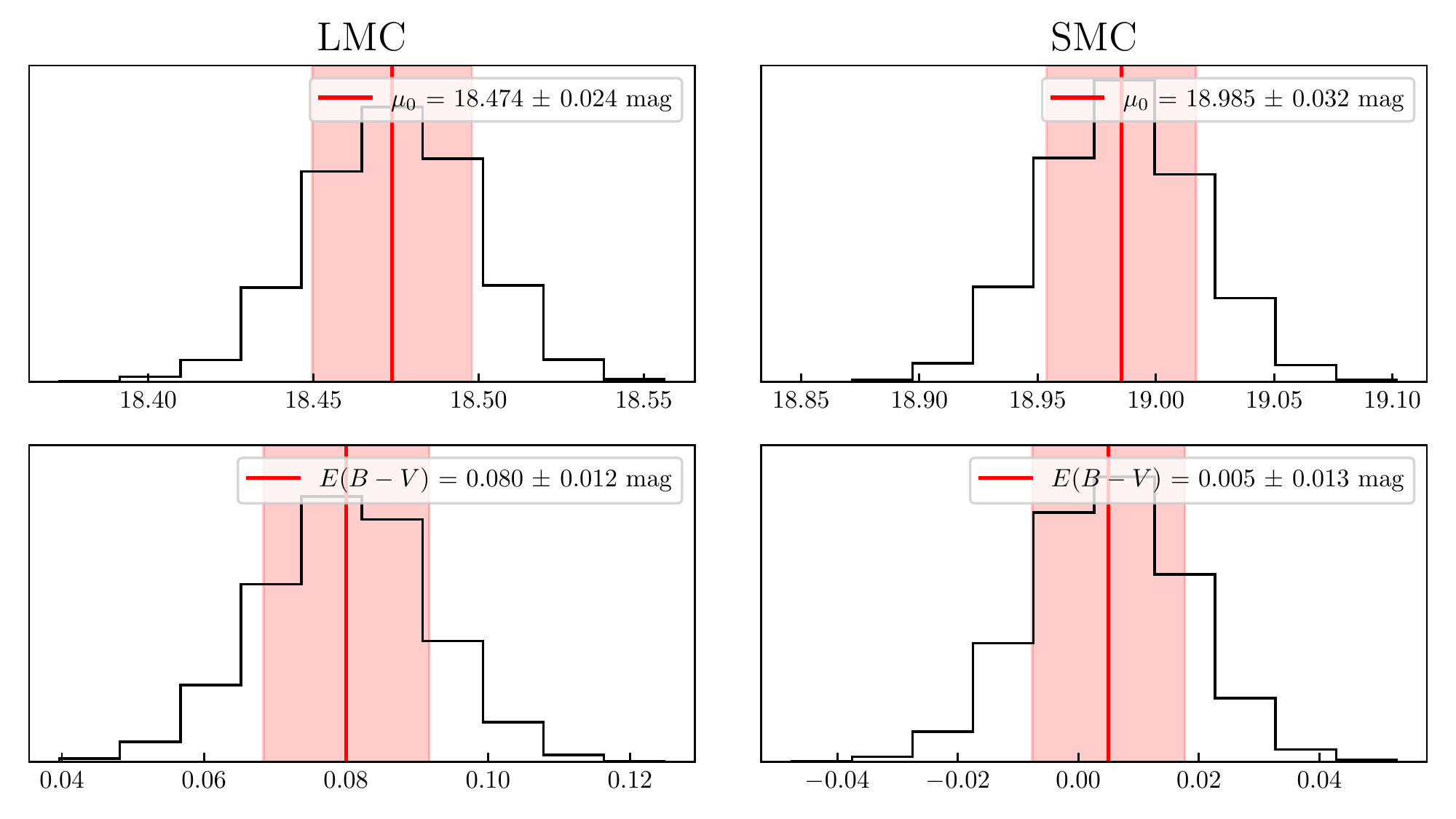}
    \caption{Distribution of distance moduli and reddenings for the bootstrapped samples. The mean values agree well with the results of the reddening law fit, though the errors are roughly an order of magnitude larger than the errors of the means. As described in the text, these larger errors likely better reflect the true statistical errors.}
    \label{fig:stat_err}
\end{figure*}

The scatter about the reddening law fit in Figure \ref{fig:final_DMs} is considerably smaller than the reported errors on the individual wavelength-dependent distance moduli. This could indicate either that the errors on the individual points are overestimated or that the scatter about the reddening curve is an underestimate of the error on the fitted true distance modulus. To discern this, we performed a bootstrapping analysis on the determination of the distance moduli and reddenings to both the LMC and SMC. 

We used random sampling with replacement to generate 10,000 homogeneous Cepheid samples in the Milky Way and SMC, each containing the same number of stars as in the real PL relations. In the LMC, the $BVI_{c}$ and longer-wavelength data sets are not homogeneous, so the process of selecting Cepheids was two-fold. To most closely mimic the data used in the reddening law fit, we generated 10,000 homogeneous samples in the $BVI_{c}$ data, and another 10,000 homogeneous samples in the $JHK_{s}$ \& $[3.6]$ data. We matched the simulated samples in these two homogeneous sets one-to-one to generate the 10,000 heterogeneous multi-wavelength samples for the LMC.

We then matched each LMC or SMC sample to a Milky Way sample and recalculated the resulting distance moduli and reddenings in each galaxy, to generate a distribution of fitted values. These distributions are shown in Figure \ref{fig:stat_err}. The calculated distance modulus error is $\pm$0.024 mag in the LMC and $\pm$0.032 mag in the SMC, which is consistent with the errors on the individual wavelength-dependent distance moduli, which average to $\pm$0.033 mag in the LMC and $\pm$0.031 mag in the SMC. We similarly find larger statistical errors on the reddenings of $\pm$0.012 mag for the LMC and $\pm$0.013 mag for the SMC. 

We speculate that these larger error values result from the correlation of individual Cepheids' locations in the PL relations across bands. Specifically, the sample sizes are sufficiently small that sample selection effects are significant. If a sample happens to have a disproportionate number of brighter Cepheids relative to the true mean, the measured mean magnitude in each band will be larger. However, this will not increase the scatter about the fit, as the effect is not random wavelength to wavelength. Thus, we conclude that the error on the mean (in Tables \ref{tab:LMC_final} \& \ref{tab:SMC_final}) are underestimates, and we adopt the larger bootstrapped error estimates as the final statistical errors on each quantity. This quantity is a reducible error, which can be decreased by a factor of $\sim \sqrt{1/n}$ by using a larger sample size.

\begin{deluxetable}{c|c|c}
\tablecaption{Irreducible EDR3 Cepheid Error Budget\label{tab:sys_bud}}
\tablecolumns{3}
\tablehead{
\colhead{Source of Uncertainty} & 
\colhead{$\sigma_{\mathrm{LMC}}$} &
\colhead{$\sigma_{\mathrm{SMC}}$}
}
\startdata
Metallicity Effects & 0.028 & 0.053\\
Zero-Point Prescription (L21b) & 0.037 & 0.037 \\
Additional ZP-offset & 0.045 & 0.045 \\
Reddening Coefficient Variation & 0.002 & 0.002 \\
\hline
\hline
Total [mag] & 0.065 & 0.079 \\
\hline
Percent Error & 3.0\% & 3.6\% \\
\enddata
\end{deluxetable}

\begin{deluxetable*}{lcccccccc}
\label{tab:met_PLRs}
\tablecaption{Fixed-Slope Period-Luminosity Relations and LMC/SMC Apparent Distance Moduli with Metallicity Corrections}

\tablehead{\colhead{Band} & \colhead{Fixed Slope} & Milky Way & \multicolumn{3}{c}{LMC} & \multicolumn{3}{c}{SMC} \\ \cmidrule(lr){3-3} \cmidrule(lr){4-6} \cmidrule(lr){7-9}
\colhead{} & \colhead{a} & \colhead{b} & \colhead{b} & \colhead{$\mu$} & \colhead{DEB error (\%)} & \colhead{b} & \colhead{$\mu$} & \colhead{DEB error (\%)}}

\startdata
$V$ & $-2.762\pm0.088$ & $-4.547\pm0.032$ & $14.133\pm0.020$ & $18.680\pm0.038$ &  & $14.328\pm0.034$ & $18.875\pm0.047$ &  \\
$I_{c}$ & $-2.987\pm0.068$ & $-5.372\pm0.028$ & $13.216\pm0.015$ & $18.589\pm0.031$ &  & $13.437\pm0.026$ & $18.809\pm0.038$ &  \\
$J$ & $-3.172\pm0.074$ & $-5.941\pm0.024$ & $12.548\pm0.017$ & $18.489\pm0.029$ &  & $12.894\pm0.030$ & $18.835\pm0.038$ &  \\
$K_{s}$ & $-3.289\pm0.056$ & $-6.391\pm0.022$ & $12.076\pm0.013$ & $18.467\pm0.026$ &  & $12.463\pm0.027$ & $18.853\pm0.035$ &  \\
\multicolumn{2}{l}{Reddening Law Fit:} & & & $\mathbf{18.430\pm0.004}$ & $-$2.1 & & $\mathbf{18.832\pm0.012}$ & $-$8.1 \\
\hline
$W_{VI}$ & $-3.355\pm0.070$ & $-6.707\pm0.023$ & $11.752\pm0.005$ & $\mathbf{18.459\pm0.024}$ & $-$0.8 & $12.033\pm0.019$ & $\mathbf{18.740\pm0.030}$ & $-$13.0 \\
$W_{JK}$ & $-3.352\pm0.051$ & $-6.718\pm0.023$ & $11.794\pm0.011$ & $\mathbf{18.512\pm0.025}$ & +1.6 & $12.123\pm0.027$ & $\mathbf{18.841\pm0.035}$ & $-$7.7 \\
\hline\hline
$V$ & $-2.762\pm0.088$ & $-4.571\pm0.032$ & $14.196\pm0.020$ & $18.767\pm0.038$ &  & $14.471\pm0.034$ & $19.042\pm0.047$ &  \\
$I_{c}$ & $-2.987\pm0.068$ & $-5.392\pm0.028$ & $13.268\pm0.015$ & $18.660\pm0.031$ &  & $13.553\pm0.026$ & $18.946\pm0.038$ &  \\
$J$ & $-3.172\pm0.074$ & $-5.949\pm0.024$ & $12.569\pm0.017$ & $18.518\pm0.029$ &  & $12.941\pm0.030$ & $18.890\pm0.038$ &  \\
$H$ & $-3.250\pm0.061$ & $-6.306\pm0.023$ & $12.205\pm0.014$ & $18.511\pm0.027$ &  & $12.614\pm0.027$ & $18.921\pm0.035$ &  \\
$K_{s}$ & $-3.289\pm0.056$ & $-6.392\pm0.022$ & $12.080\pm0.013$ & $18.472\pm0.026$ &  & $12.471\pm0.027$ & $18.863\pm0.035$ &  \\
\multicolumn{2}{l}{Reddening Law Fit:} & & & $\mathbf{18.437\pm0.005}$ & $-$1.8 & & $\mathbf{18.853\pm0.009}$ & $-$7.0 \\
\hline
$W_{VI}$ & $-3.355\pm0.070$ & $-6.718\pm0.023$ & $11.780\pm0.005$ & $\mathbf{18.498\pm0.024}$ & +1.0 & $12.096\pm0.019$ & $\mathbf{18.814\pm0.030}$ & $-$9.1 \\
$W_{JK}$ & $-3.352\pm0.051$ & $-6.719\pm0.023$ & $11.796\pm0.011$ & $\mathbf{18.515\pm0.025}$ & +1.8 & $12.128\pm0.027$ & $\mathbf{18.847\pm0.035}$ & $-$7.3 \\
\enddata
\tablecomments{Values above the double horizontal line are using metallicity corrections from \cite{2018AA...620A..99G}, and values below are using corrections from \cite{2021arXiv210310894B}. Quantities in \textbf{bold} indicate final $\mu_{0}$ values.}
\end{deluxetable*}

\subsection{Metallicity}\label{subsec:sys_met}

Unfortunately, the effect of metallicity on Cepheid luminosity, particularly its effect at specific wavelengths, is still a subject that is being actively debated. Stellar models of Cepheids do not produce consistent values of the metallicity effect, though there is a substantial body of literature pointing to empirically-determined metallicity effects in the optical of around $-0.2$ mag/dex \citep[e.g.][]{1998ApJ...498..181K, 2001ApJ...553...47F, 2004ApJ...608...42S, 2004A&A...415..531S, 2006ApJ...652.1133M, 2009MNRAS.396.1287S, 2016ApJ...826...56R, 2021arXiv210310894B}. However, many of these studies have error bars which are nearly as large as the quoted magnitude of the metallicity effect itself, and even very recent determinations based on \textit{Gaia} EDR3 data have disagreed by a factor of two \citep{2021arXiv210310894B, 2021arxiv210811391R}. Further, some studies \citep[e.g.][]{2001AcA....51..221U, 2017ApJ...842..116W} find the metallicity effect to be nearly zero, or even to have a positive coefficient \citep{2008A&A...488..731R}. We have not applied metallicity corrections to the main distance modulus fits given in Tables \ref{tab:LMC_final} and \ref{tab:SMC_final}, but we test the effects of metallicity corrections, as described below.

To test the effects of metallicity corrections on our results, we apply the corrections from \citep{2018AA...620A..99G, 2021arXiv210310894B} (G18, B21) to all available bandpasses. This includes $VI_{c}JK_{s}$ and $W_{VI}$ \& $W_{JK}$ in G18, and additionally includes the $H$ band in B21. The results of these corrections are given in Table \ref{tab:met_PLRs}. For the optical and near-infrared bandpasses, we performed a reddening curve fit to determine the distance moduli, as in the main analysis. Note that both sets of reddening corrections perform quite well for the LMC, resulting in errors on the level of 1-2\% compared to the DEB distances. However, the corrections uniformly underestimate the distance modulus of the SMC by several percent, indicating that they are \textit{overcorrecting} the metallicity term. This could be evidence that the discrepency with DEB distances is not due to a strong metallicity effect, but is rather due to an average parallax offset for the Milky Way Cepheids. However, with the data currently available to us, these effects are not easily distinguishable due to uncertainties on the metallicities of individual Milky Way Cepheids and the individual parallaxes.

Based on the current data, we now use an external constraint to place bounds on the metallicity effect. We assume that the true differential distance modulus between the SMC and LMC is equal to the difference in detached eclipsing binary (DEB) distance moduli, given that the DEB determinations are geometric and should be relatively insensitive to differential metallicity effects. In the case of the SMC and the LMC, the DEB differential distance modulus is $\mu_{\mathrm{SMC}} - \mu_{\mathrm{LMC}} = (18.977\pm0.044) - (18.477\pm0.030) = 0.500\pm0.056$ mag. Our final differential distance modulus (derived from Tables \ref{tab:LMC_final} \& \ref{tab:SMC_final}) is 18.983 - 18.472 = 0.511 mag, well within the $\pm$0.056~mag error of the differential modulus derived from the DEBs. Moreover, we can use this quantity to put a bound on the magnitude of the metallicity effect. 

The metallicity of the LMC is [Fe/H]$=-0.33 \pm 0.03$ dex and of the SMC is [Fe/H]$=-0.75 \pm 0.02$ dex \citep{2008A&A...488..731R}. The Milky Way Cepheids have a mean metallicity of $+0.13 \pm 0.02$ dex according to the catalog of \cite{2018A&A...619A...8G}. The difference in metallicity between the LMC and Milky Way is then 0.46 dex; between the SMC and Milky Way it is 0.88 dex, and between the LMC and SMC it is 0.42 dex. Using a simple interpolation of the 0.011 mag difference in the differential distance modulus between the LMC and SMC, we obtain potential metallicity effects of $\sigma_{[Fe/H]} = 0.46\times(0.011 / 0.42) = 0.012$ mag in the LMC and $\sigma_{[Fe/H]} = 0.88\times(0.011 / 0.42) = 0.023$ mag in the SMC. Since these effects are well below the level of the errors on the above calculations, we will adopt the systematic error floor instead, as calculated below.

There is an uncertainty on the effect of metallicity for any given Cepheid sample. This error cannot be reduced by increasing the sample size or constructing a more robust sample, and the magnitude of the effect depends on the metallicity difference between the calibrating sample and the sample to which a distance is being measured. As mentioned above, the exact effect of metallicity on the PL relations is still highly uncertain. Even disregarding the differences between the values found by different studies, the corrections are reported to have uncertainties between 0.19 and 0.05 mag/dex in G18 and 0.05 and 0.09 mag/dex in B21. The Wesenheit functions tend to have the lowest uncertainties on their metallicity calibrations, averaging 0.06 mag/dex. This results in a \textit{minimum} systematic floor of 0.06 mag/dex in the metallicity effect when using mean metallicities for each galaxy. Propagated to their metallicity differences, this would be a systematic uncertainty of 0.06 $\times$ 0.46 = 0.028 mag in the LMC and 0.06 $\times$ 0.88 = 0.053 mag in the SMC. We adopt these quantities as the systematic error due to metallicity for our measurement, given in Table \ref{tab:budget}.

\subsection{Zero-point Correction and Offset\label{subsec:sys_zpL20}}

\cite{2020arXiv201203380L} quote a minimum $\pm$10 $\mu$as root-mean square variation of the parallax systematics on large angular scales. We propagate this uncertainty by recalculating the distance moduli and reddenings to both galaxies using offsets of $+10~\mu\textrm{as}$ and $-10~\mu\textrm{as}$. The resulting uncertainties are $\pm0.037$ mag on the distance moduli and $\pm0.001$ mag on the reddenings. This is an irreducible systematic error resulting from the current uncertainties on the \textit{Gaia} parallaxes.

Further, the zero-point offset calculated in Section \ref{subsec:DEB_comp} carries an uncertainty of $\pm$14 $\mu$as. We propagate this uncertainty similarly using offsets of $+18 - 14 = +4~\mu\textrm{as}$ and $+18 + 14 = +32~\mu\textrm{as}$. The resulting uncertainties are $\pm0.054$ mag on the distance moduli and $\pm0.002$ mag on the reddenings, as given in Table \ref{tab:budget}. This error is a systematic particular to our analysis and sample. However, as it stands, the introduction of the additional parallax offset is a significant uncertainty for all Cepheid samples. With studies using bright stars finding significant parallax offsets between about $-17$ $\mu$as (R21) and $+7$ $\mu$as \citep{2021arXiv210110206M}, the uncertainty in the field as a whole appears to be about $\pm12$ $\mu$as, which results in an uncertainty of $\pm0.045$ mag. However, the metallicity correction is covariant with this uncertainty; using a smaller metallicity correction is similar to using a more negative parallax offset term. We include both terms separately in the error budget in Table \ref{tab:sys_bud}, but we note that these errors can correlate.

\subsection{Reddening Corrections\label{subsec:sys_red}}

We investigate uncertainties due to the adjustment of the reddenings to the updated system of \cite{2016RMxAA..52..223T}. Then, we address the irreducible systematic uncertainty due to differences in the reddening coefficients between different galaxies.

\subsubsection{Reddening System Adjustment}\label{subsubsec:red_adj}

In Section \ref{subsec:Gaia_MW_dists}, we updated the reddenings from \cite{2007AA...476...73F} (F07) to the newer system of \cite{2016RMxAA..52..223T}. This adjustment was performed by multiplying all reddenings by a factor of $1.055 \pm 0.034$. To propagate the uncertainty on this quantity, we recalculate the distance moduli and reddenings to the LMC and SMC after multiplying the F07 Milky Way reddenings by factors of $1.055 - 0.034 = 1.021$ and $1.055 + 0.034 = 1.089$. The resulting errors on the distance moduli are below their level of precision ($<0.0005$ mag), and the errors on the reddenings are $\pm0.014$ mag. This is a reducible systematic particular to our sample.

\subsubsection{Reddening Coefficient Variance}\label{subsubsec:sys_red_coeff}

It is standard practice to assume that the reddening coefficient $R_{V}$ is constant from galaxy to galaxy; however, some studies have indicated this may not be the case \cite[e.g.][]{2003ApJ...594..279G}, and $R_{V}$ has been shown to vary even within the Milky Way \citep[e.g.][]{1999PASP..111...63F, 2013ApJ...769...88N}. \cite{2021arXiv210609400M} investigated the effect of allowing $R_{V}$ to vary by galaxy, and they found that it could have a significant impact on the value of the Hubble constant due to a systematic difference between the reddening coefficients in the supernova calibrating galaxies and in the supernova galaxies that extend into the Hubble flow.

Fortunately, the differences in mean $R_{V}$ between the Milky Way and the Magellanic Clouds are estimated to be relatively small. The mean Milky Way $R_{V}=3.1$ \citep{1989ApJ...345..245C}, while the mean LMC $R_{V}=3.4$ \citep{2003ApJ...594..279G}, and the mean SMC $R_{V} = 2.7$ \citep{1985A&A...149..330B}. This results in a systematic difference in distance moduli of only 0.002 mag from the reddening curve fit, and $<0.006$ mag difference on the fitted reddenings. However, the effect on the Wesenheit distance moduli can be significant, with the LMC changing by $-0.09$ mag and the SMC changing by $+0.14$ mag in $W_{VI}$. The effect in the near-infrared Wesenheit function $W_{JK}$ is $<0.001$ in both galaxies. In the SMC, using the three-band Wesenheit function $W_{H,\,VI}$ changes the resulting distance modulus by $+0.05$ mag. This emphasizes the sensitivity of the optical-color Wesenheit measurements to differences in the assumed $R_{V}$. In contrast, a reddening law fit remains remarkably stable, since the reddening is actually fit from the data rather than assuming a specific numerical relationship between color and reddening. Since we are able to perform a reddening law fit in the calibrator galaxies, we will adopt the much smaller error of 0.002 mag as the irreducible systematic error due to reddening coefficient differences. However, this may be a significant issue for the calculation of $H_{0}$ in more distant galaxies, as discussed by \cite{2021arXiv210609400M}.

\subsection{Magellanic Clouds Geometry}\label{subsec:sys_geom} 

A small error may be introduced into the distance measurements due to the geometry of the Magellanic Clouds. The LMC has a slight tilt which results in distribution of distances across the face of the galaxy of $\pm1.5$ kpc; the more complicated geometry of the SMC has a wider distribution of distance $\pm5$ kpc \citep{2021arXiv210310894B}. If the Cepheids happen to lie disproportionately on closer or farther regions of the galaxy, this can bias the distance measurement. The scatter due to this effect is fully accounted for in the bootstrapping error estimate; however, we also constrain the potential systematic effects here.

As in the analysis of \cite{2021arXiv210310894B}, we calculate the Cartesian distances of Cepheids from the center of each of the two galaxies, and then use the distance formulae of \cite{2016AcA....66..149J} in the LMC and \cite{2020ApJ...904...13G} in the SMC to determine deviations from mean distance for each Cepheid. We find that the average difference from the mean is 0.118 kpc in the LMC and $-0.721$ kpc in the SMC. This is a difference in distance modulus of $0.005$ mag in the LMC and $-0.025$ mag in the SMC. We adopt these full quantities as the systematic errors on each measurement. We categorize this as a particular systematic of our sample since, in principle, having a larger sample which is better distributed across the faces of the galaxies would reduce this error.

We finally note that if metallicity corrections are applied as in Section \ref{subsec:sys_met}, using these quantities to correct the distance moduli would tend to increase the discrepancies between the distances according to DEBs. If metallicity corrections are not applied, it would modestly decrease the calculated value of the additional parallax offset to $+16$ $\mu$as. Regardless, accounting for this geometry cannot reconcile the distances to the Magellanic Clouds, so we choose to adopt the quantity as a systematic error.

\section{Summary and Conclusions}\label{sec:summary}

We have explored the errors underlying the Cepheid period-luminosity relation as derived from \textit{Gaia} EDR3 parallaxes. We find that there are significant irreducible systematics, resulting both from the uncertainties on EDR3 parallaxes for bright stars and from the covariance of the multi-wavelength effect of metallicity with any additional derived parallax offset term. At this time, we estimate the absolute minimum error on local EDR3-based Cepheid calibrations is at the level of 3\%. Combined with the statistical uncertainties and systematic uncertainties associated with a particular sample, this will result in calibration uncertainties at least at the 4\% level. We note that using Wesenheit magnitudes rather than a multi-wavelength reddening law fit will tend to increase this number due to potential differences in the reddening coefficients between galaxies. We find that the best overall fit to the existing data uses metallicity corrections that are consistent with zero and an additional global parallax offset of $+18$ $\mu$as. However, due to the strong covariance of metallicity and the additional parallax offset, we do not advocate for this quantity as a universal term. Using this fit, for our sample of 37 Galactic Cepheids with broad wavelength coverage, the Gaia parallaxes provide a distance to the LMC precise to 4.4\% and to the SMC precise to 5.5\%. We look forward to a more precise and accurate Galactic calibration of the Cepheid PL relation with future data releases.

\section{Acknowledgements}

We thank the University of Chicago and the Carnegie Institution for Science for their continuing generous support of our long-term research into the expansion rate of the Universe.  

This work has made use of data from the European Space Agency (ESA) mission \textit{Gaia} (\url{https://www.cosmos.esa.int/gaia}), processed by the \textit{Gaia} Data Processing and Analysis Consortium (DPAC, \url{https://www.cosmos.esa.int/web/gaia/dpac/consortium}). Funding for the DPAC has been provided by national institutions, in particular the institutions participating in the \textit{Gaia} Multilateral Agreement. 

This research was supported by NASA/HST grant AR-16126 from the Space Telescope Science Institute, which is operated by the Association of Universities for Research in Astronomy, Inc., under NASA contract NAS5-26555.

\software{Astropy \citep{2013A&A...558A..33A, 2018AJ....156..123A}, NumPy \citep{2011CSE....13b..22V}, Matplotlib \citep{2007CSE.....9...90H}, scipy \citep{2020NatMe..17..261V}}

\facility{Gaia}

\appendix

\section{Detailed Milky Way Sample Refinement}\label{sec:A_refinement}

From the total sample of 59 Milky Way Cepheids, 22 have been removed from the analysis presented in the main body of this paper: 8 Cepheids were automatically removed by period cuts; 11 Cepheids were removed by a cut on the goodness of fit parameter {\tt ruwe}; \textbf{Y Oph} is suspected of being an overtone pulsator; \textbf{SU Cru} was eliminated because of its high fractional-parallax error; and finally, \textbf{SV Vul} was removed due to very large residuals in the long-wavelength PL relations.

\subsection{Period Cuts}

We adopted a lower period cut of 5 days ($\log P > 0.7)$, removing from the sample the following short period Cepheids: \\
\textbf{FF Aql}, \textbf{QZ Nor}, \textbf{RT Aur}, \textbf{SU Cyg}, \textbf{T Vel}, \textbf{T Vul}, \textbf{VZ Cyg}, \textbf{Y Lac}

\smallskip

We additionally calculated fits for a lower period cut of 10 days ($\log P > 1.0$) in Tables \ref{tab:LMC_final} and \ref{tab:SMC_final}, which removes the following Cepheids which otherwise pass quality cuts: \\
\textbf{BG Lac}, \textbf{CV Mon}, \textbf{V Cen}, \textbf{Y Sgr}, \textbf{CS Vel}, \textbf{BB Sgr}, \textbf{V Car}, \textbf{U Sgr}, \textbf{V496 Aql}, \textbf{X Sgr}, \textbf{GH Lup}, \textbf{S Nor}

\subsection{EDR3 Fit Quality Cuts}

\begin{figure}
    \figurenum{A1}
    \centering
    \includegraphics[width=280pt]{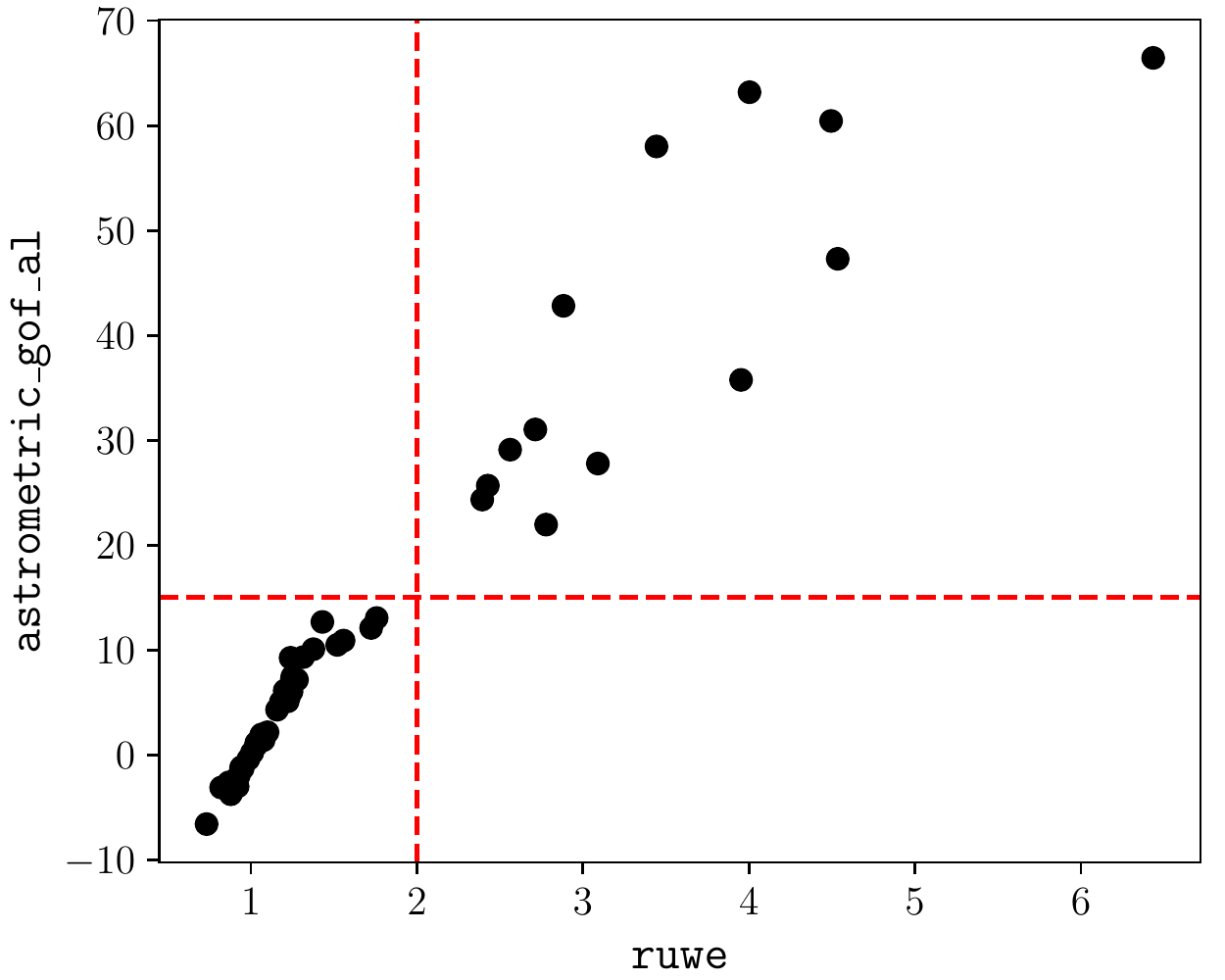}
    \caption{Astrometric goodness of fit ({\tt astrometric\_gof\_al}) versus unit weight uncertainty ({\tt ruwe}) for all Cepheids in the Milky Way sample. We perform an upper cut of 2.0 on {\tt ruwe}, which is equivalent to a cut {\tt astrometric\_gof\_al} $\lesssim14.0$ for our particular sample.}
    \label{fig:gof_ruwe}
\end{figure}

The goodness-of-fit parameter {\tt ruwe} is recommended as a primary indicator of poor astrometric solutions in \cite{2020arXiv201203380L}. Thus, we use a cut {\tt ruwe} $\leq 2.0$, which is equivalent to {\tt astrometric\_gof\_al} $\lesssim 14.0$, as shown in Figure \ref{fig:gof_ruwe}. This is similar to the cut of \cite{2021ApJ...908L...6R} who use {\tt astrometric\_gof\_al}$\leq12.5$, and we find that using their value of 12.5 would remove only two more Cepheids and would not impact the final results. We note that \cite{2021arXiv210110206M} recommend using a cut {\tt ruwe}$<1.4$; however, using a cut upwards of 2.0 is safe, although the reported parallax errors become increasingly underestimated. A more stringent cut of {\tt ruwe}$<1.4$ did not significantly change the outcome of our fits, although the statistical errors increase due to the diminished sample size. This cut removes the following Cepheids:\\
\textbf{V350 Sgr}, \textbf{$\bm{\delta}$ Cep}, \textbf{U Aql}, \textbf{$\bm{\eta}$ Aql}, \textbf{W Sgr}, \textbf{U Vul}, \textbf{S Sge}, \textbf{S Mus}, \textbf{$\bm{\beta}$ Dor}, \textbf{$\bm{\zeta}$ Gem}, \textbf{l Car}
 
\subsection{Overtone Pulsators}

Overtone Cepheids are a subclass of Cepheid variable stars. The majority of Classical Cepheids are categorized as ``fundamental pulsators," and it is these stars that are generally used to determine distances via the Leavitt Law. Overtone pulsators are hotter Cepheids with their ionization zones closer to the surface than their fundamental counterparts. The periodic heating and expansion of Cepheids is driven primarily by increasing opacity of the \ion{He}{2} partial ionization zone upon compression. In a fundamental pulsator, this zone is located deep within the star and expansion is unidirectional; however, in overtone pulsators, the partial ionization zone lies sufficiently close to the surface to cause the radial gas flow to resonate in a higher-order harmonic. This reduces the observed period by as much as 50\% \citep{1980PASP...92..165C}. As these Cepheids have similar luminosities but shorter periods than fundamental Cepheids, they will appear to lie above the canonical period-luminosity relations and need to be removed from PL fitting to avoid biasing the intercept to brighter values.

We remove \textbf{Y Oph} since it is categorized in the {\it General Catalogue of Variable Stars} \citep{2017ARep...61...80S} as ``DCEPS," which are Cepheid variables having low amplitudes and almost symmetrical light curves, making them likely to be overtone pulsators.

\subsection{Fractional Parallax Error}

\cite{2020arXiv201206242F} recommend excluding stars having high ratios of parallax error to parallax ($\sigma_{\pi} / \pi$). All but one of the stars in our sample had a fractional parallax error below 0.1, while $\sigma_{\pi} / \pi$ for \textbf{SU Cru} was $0.145 / 0.159 = 0.915$, well outside of the typical range. It is evident that the parallax is compromised, as SU Cru lies $>10\sigma$ from the PL fits.

\subsection{PL Relation Outlier}

We exclude \textbf{SV Vul} because it is a significant outlier in the long-wavelength PL relations. Specifically, it is a $>4\sigma$ outlier in the $[3.6]$ and $[4.5]$ bands, and a $>2.5\sigma$ outlier in the $JHK_{s}$ bands. Notably, the residual between its magnitude in each band and the ridge of the PL relation does not decrease to the degree expected if this scatter were due to the Cepheid lying on the extreme blue end of the instability strip. The difference between the largest residual (in the $B$ band) and the smallest residual (in the $[3.6]$ band) is only 0.087 mag, while the residual in the $[3.6]$ band is 0.388 mag. This indicates the large residuals are likely caused by a $\gtrsim0.3$ mag distance error. For this reason, we exclude SV Vul from our sample, despite not having direct evidence from the \textit{Gaia} parallaxes or quality metrics. Finally, we also note that SV Vul was previously identified as an outlier in \cite{2021ApJ...908L...6R}, who used a different set of photometry, indicating that this is more likely to be a parallax-based error, rather than intrinsic scatter or photometric error.

\end{document}